\documentclass{biophys-new}
\usepackage[utf8]{inputenc}
\usepackage{graphicx}
\usepackage[colorlinks,allcolors=cyan!70!black]{hyperref}

\usepackage{subfigure}

\usepackage{soul}

\usepackage{lipsum}

\title{Neuron Signal Propagation Analysis of Cytokine-Storm induced Demyelination}
\runningtitle{Neuron Signal Propagation Analysis of Cytokine-Storm induced Demyelination} 

\author[1,*]{Geoflly L. Adonias}
\author[2]{Harun Siljak}
\author[3]{Michael Taynnan Barros}
\author[1]{Sasitharan Balasubramaniam}
\runningauthor{G. Adonias et al.} 

\affil[1]{Telecommunications Software and Systems Group, Waterford Institute of Technology, Waterford, Ireland.}
\affil[2]{Department of Electronic and Electrical Engineering, Trinity College Dublin, Dublin, Ireland.}
\affil[3]{School of Computer Science and Electronic Engineering, University of Essex, Colchester, United Kingdom.}

\corrauthor[*]{gadonias@tssg.org}

\papertype{Article}

\begin{document}

\begin{frontmatter}

\begin{abstract}
The COVID-19 pandemic has shaken the world unprecedentedly, where it has affected the vast global population both socially and economically. The pandemic has also opened our eyes to the many threats that novel virus infections can pose for humanity. While numerous unknowns are being investigated in terms of the distributed damage that the virus can do to the human body, recent studies have also shown that the infection can lead to lifelong sequelae that could affect other parts of the body, and one example is the brain. As part of this work, we investigate how viral infection can affect the brain by modelling and simulating a neuron's behaviour under demyelination that is affected by the cytokine storm. We quantify the effects of cytokine-induced demyelination on the propagation of action potential signals within a neuron. We used information and communication theory analysis on the signal propagated through the axonal pathway under different intensity levels of demyelination to analyse these effects. Our simulations demonstrate that virus-induced degeneration can play a role in the signal power and spiking rate and the probability of releasing neurotransmitters and compromising the propagation and processing of information between the neurons. We also propose a transfer function that models these attenuation effects that degenerates the action potential, where this model has the potential to be used as a framework for the analysis of virus-induced neurodegeneration that can pave the way to improved understanding of virus-induced demyelination. 
\end{abstract}

\begin{sigstatement}
Virus infections are increasingly linked to neurological pathologies as these viral agents also show neurotropic properties. Despite their success in invading the brain, neurotropic viruses can cause pro-inflammatory immune responses that can last for months after the virus is cleared from the nervous system. Several studies have shown that this neuroinflammation can lead to neurodegeneration. Unfortunately, a neuron that has undergone a demyelinating degeneration never comes back to its healthy state. The present study provided a quantitative analysis concerning cytokine-induced demyelination triggered by a viral infection. As such, it provides a model that describes how signal propagation and neuronal communication is compromised. It also offers a framework for the study of demyelination as an attenuating factor.
\end{sigstatement}
\end{frontmatter}

\section*{Introduction}
\label{sec:introduction}

The recent outbreak of the \textbf{coronavirus disease 2019} (COVID-19) pandemic caused by the \textbf{severe acute respiratory syndrome coronavirus 2} (SARS-CoV-2) has been impacting society as a whole, leaving a long-lasting impact on people's health. 
As a result, the scientific community have been joining efforts and resources towards 
not only the epidemiologic characteristics and transmission dynamics of the virus, but also the physiological damage to the human body as the infection is prolonged. SARS-CoV-2 is well known for affecting primarily the respiratory system and can potentially leave lifelong sequelae in tissues and organs. Furthermore, there has been an increase in works that suggest SARS-CoV-2 may be able to invade the nervous system~\cite{Zubair2020, SanclementeAlaman2020, MONTALVAN2020, Song2020} and elicit neurodegeneration~\cite{Zanin2020, WU2020}, in which sequelae may have other detrimental effects on patients' lives post-infection.

\begin{figure}[t]
    \centering
    \includegraphics[width=.85\textwidth]{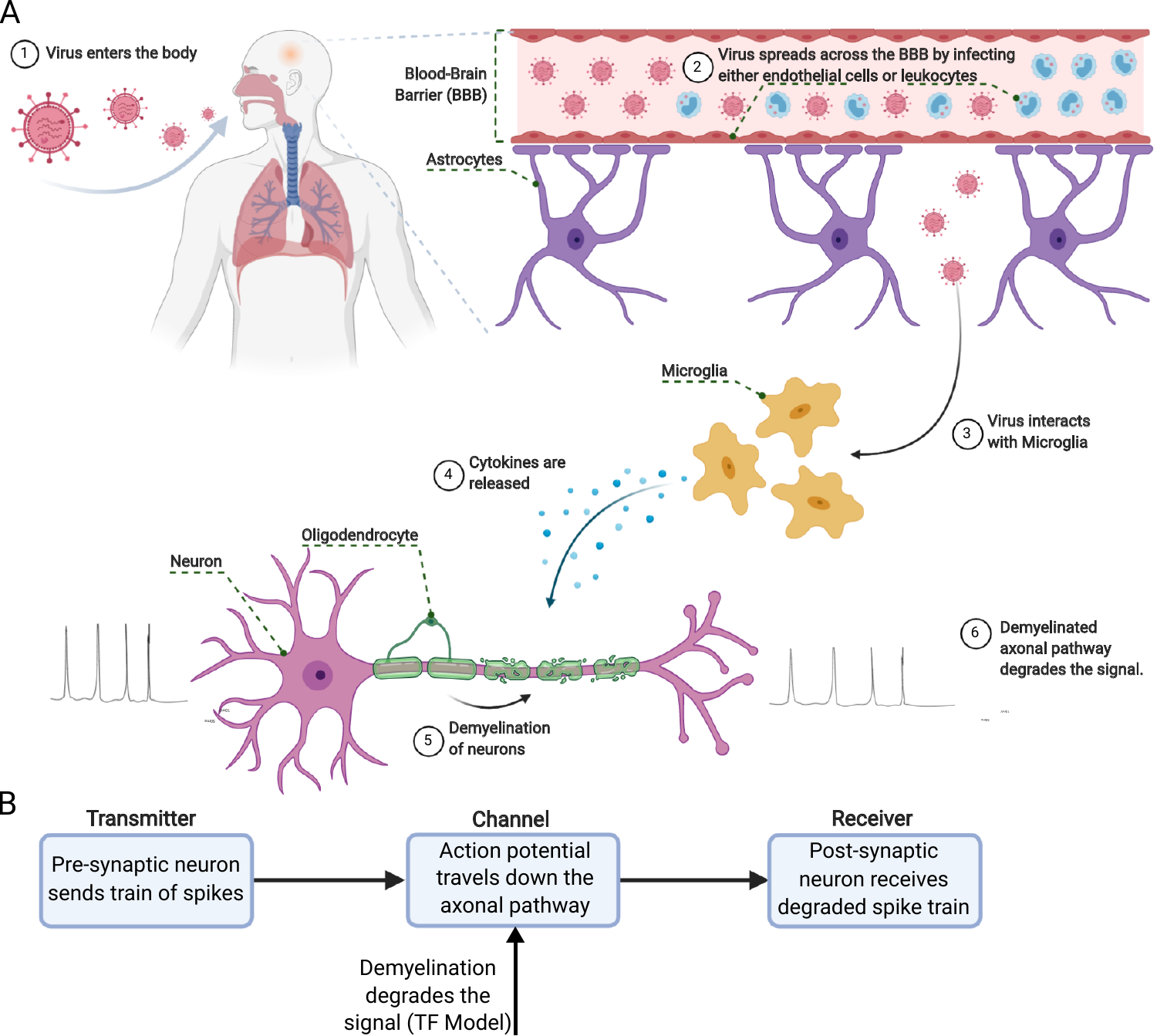}
    \caption{Pathogenesis of a virus-induced demyelination.}
    \label{fig:viral_persistence_demyelination}
\end{figure}

Viruses that present the ability to infect nerve cells are known to exhibit \textit{neurotropic} properties and can also be called \textit{neuroinvasive}. By infecting cells in the nervous system and replicating themselves within it, these viruses can negatively impact neurological functions and even cause severe nerve damage {\color{black}by triggering a pro-inflammatory immune response}  \cite{WU2020}. 
Unfortunately, SARS-CoV-2 is not the only virus that exhibits this kind of behaviour. For example, it has been shown that the Zika virus (ZKV)~\cite{Oh2017} can infect the peripheral nervous system (PNS) and, sometimes, spread to the central nervous system (CNS). Furthermore, viruses such as the human immunodeficiency virus (HIV)~\cite{Valcour2012}, can infect the CNS and trigger \textbf{immunotherapy-induced neuroinflammation} and, consequently, lead to neurodegeneration~\cite{Cheng2018}. 
Viruses are known for causing dramatic structural and biochemical changes to the {\color{black}host cell, by hijacking and exhausting its machinery for replication until, eventually, the cell is killed.}  {\color{black}This viral manipulation of a host cell}  provokes neuroinflammatory defence mechanisms that can be characterised by numerous toxic-metabolic derangements, such as cytokine storms
(Figure~\ref{fig:viral_persistence_demyelination}(a)). Such inflammation could potentially lead to several types of neurodegeneration, including \textbf{demyelination}. {\color{black}As cytokines are released to fight the infection, healthy tissues could be affected as a ``collateral damage'' of the fight against the infectious agent}~\cite{Stohlman2001}.

Other types of coronavirus have been known to cause demyelination, such as the murine coronavirus (M-CoV). It has been identified to cause demyelinating disease and even after the virus is cleared from the CNS, the demyelination can continue for a few months~\cite{Stohlman2001}.  This behaviour also matches findings on SARS-CoV-1, which reports a decrease in viral titers as clinical disease worsens~\cite{Dandekar2005}. M-CoV is a type coronavirus of the same genus (\textit{betacoronavirus}) as SARS-CoV-2, and it is believed to be $43.8\%$-$48\%$ similar to this novel coronavirus~\cite{Geldenhuys2018}. SARS-CoV-2, when compared to SARS-CoV-1, triggers lower levels of interferons and pro-inflammatory cytokines and chemokines. However, it is capable of infecting and replicating a significantly higher amount of virus in human tissues~\cite{Chu2020}.

The virus replication process produces host cells exhaustion that leads to the activation of the immune system, which calls for macrophages' work. Those are types of cells of the immune system, and their resident in the brain is the microglia~\cite{Zubair2020, Song2020}. Macrophages secret pro-inflammatory cytokines such as interleukin-1 (IL-1), interleukin-6 (IL-6) and tumor necrosis factor \textit{alpha} (TNF-$\alpha$) aiming to fight the infection. These cytokines exert cytotoxic effects on neurons and glial cells, e.g. oligodendrocytes, where this damages the neurons' myelin sheath providing support and insulation to axons~\cite{Merson2010}.

Then the dynamics of cytokine storms and their efficiency in fighting infections without further cellular damage becomes of increase attention to measure the neuroinflammatory effects of COVID-19 infection. Ludwig et al~\cite{Ludwig1268} found out that as the neuropathy is more severe, the higher the serum concentration of TNF-$\alpha$ and IL-6. This matches more recent results found by Chen et al~\cite{Chen2020}, which analysed plasma cytokine levels concerning the severity of COVID-19 cases. They identified a positive correlation {\color{black} between the levels of cytokines and the severity of the COVID-19 cases,} meaning that the more severe the COVID-19 cases were, the higher were the levels of cytokines. On the other hand, the dynamics of a cytokine storm can be similar for different inflammatory scenarios, yielding the need for a more general approach that evaluates cytokine dynamics.
An example is the work of Waito et al~\cite{Waito2016} in which they match recordings of 13 different cytokines with a model of non-linear ordinary differential equations. Likewise, Yiu et al~\cite{Yiu2012} analysed the dynamics of cytokine storms and provided evidence for how cytokines induce or inhibit other cytokines. Additionally, some pieces of literature report the effects of cytokine storms on specific neuronal structures or non-neuronal cells. For instance, the work of Bitsch et al~\cite{Bitsch2000} shows a negatively correlated relationship between the amount of microglia-produced TNF-$\alpha$ and the concentration of myelin oligodendrocyte glycoprotein (MOG). It shows that the more TNF-$\alpha$ there is, the less MOG oligodendrocytes will produce, compromising the myelin sheath structure.
On the other hand, Redford et al~\cite{Redford1995} showed how the number of axons found in sciatic nerves is affected. They found that the more the concentration of TNF-$\alpha$ is increased, the more axons are found to be damaged. However, the complete biophysical models of these various effects to neurons caused by infection, particularly infections from COVID-19, require urgent attention since correct treatment procedures for acute infection damage can benefit from mathematical modelling.

In the past decade, Molecular Communications (MC) has been improving biological models by accounting for the communication of cells using their signalling mechanisms as information carriers. MC bridges electrical and communications engineering, molecular biology and biomedical engineering and provides complete end-end models of biophysical transmission of molecules, their propagation, and their reception. 
A recent survey~\cite{barros2020molecular} reports numerous works concerning the use of MC for the analysis and modelling of infectious diseases.  However, accounts for the effects of infections are still missing from a biophysical approach even within MC models, as the COVID-19 effects are many, and the molecular interactions with the body can be used to predict the behaviour of a population of cells, even tissues and organs.
The emergence of novel MC models for biophysical processes, such as the demyelination induced by COVID-19, delivers an in-depth analysis of tissue behaviour that is needed for synthetic biology-based treatment. Even further targeted drug delivery technology can alleviate the cytokine storms effects on neurons and provoke the restoration of the myelin sheath~\cite{Chahibi2017}.

This paper presents a systems theory-based analysis of the demyelination that is, either directly or indirectly, caused by viral infections from the perspective of MC, more specifically, a neuro-spike MC. The goals of this paper are (1) to propose a mathematical model that represents the cytokine storms stemming from acute infection-derived inflammation; (2) to provide insights on how the demyelination will affect the neuronal information along the axonal pathway; and (3) to provide a transfer function (TF) model that describes the effects of the membrane action potential caused by the degeneration of the myelin sheath. {\color{black}This TF can be considered fundamentally as a model of the demyelination process itself. It is intrinsically tied with the behaviour of equivalent resistance-capacitance (RC) circuits, but at the same time, has a reduced complexity as it opts for exponential asymptotics. Given that complex cascades of equivalent simple blocks (in our example, myelin sheaths) are traditionally well-approximated by first-order transfer functions with time delay and the underlying physical rationale, our model joins the family of established, applicable biophysical transfer function models.} We expect that these analyses could pave the way for more in-depth studies that can support \textit{in vitro} and \textit{in vivo} experimental work on the neurological effects induced by neurotropic viral infections.

The contributions of this work are as follows:

    \begin{itemize}
        \item \textit{A mathematical model that describes the process of virus-induced demyelination:} We present a model starting from the well-documented fact that viral infections trigger cytokine storms. These storms are counter-measures of the immune system against the infection. We investigate the signal power, the signal attenuation and magnitude-squared coherence (MSC) analysis of the axonal pathway as a communication channel. 
        
        \item \textit{An analysis of the effects of demyelination on the neuronal action potential propagation:} We conduct a variety of {\color{black} analyses, such as latency, attenuation and spiking rate},  on the spike trains that pass through a demyelinated pathway. Furthermore, we also analyse how the intensity of cytokines storms correlates with the amount of attenuation present in the signal at the output of the neuron.
        
        \item \textit{A transfer function model that accounts for the effects of the demyelination-induced attenuation:} We propose a transfer function that describes the transition of a healthy to an unhealthy neuron. The model accounts for sheath-by-sheath demyelination which affects membrane potential, peak times and spike width. This should lead to more in-depth analysis and open a new view on demyelination modelling, especially those triggered by a neurotropic viral infection and how it affects neurons' intra-signalling.
    \end{itemize}


\section*{Materials and Methods}
\label{sec:mat_met}

Models that describe the dynamics and evolution of cytokine concentrations, without regard to the cells that secrete or are affected by them, have been proposed by Yiu et al~\cite{Yiu2012}. This model was extended by coupling together with a neuronal model that implements the behaviour of a myelinated axon~\cite{Cohen2020}. {\color{black}By extending these models, we developed simulations and the default parameters of the cell  (such as the length and diameter of each of their compartments) were based on the original model. Therefore, the myelin sheath properties concerning their morphological characteristics were not modified or changed. We studied and quantified the propagation of action potentials on a Layer 5 (L5) pyramidal cell with its myelin sheath affected. All simulations were performed with extensions to the NEURON Simulator~\cite{Carnevale2009}.} 

\subsection*{Cytokine Signalling in Microglia}
\label{subsec:cyt_sig_glia}

The growth and decay of an individual cytokine's response to its given initial state are first represented by a second-order, linear, time-invariant ordinary differential equation. Denoting the serum concentration, $\rho(t)$, and its rate of change, $\Delta_{\rho}(t)$, in a vector-matrix form, can be represented as follows

\begin{equation}
    \begin{bmatrix}
        \dot{\rho}(t) \\
        \dot{\Delta_{\rho}}(t)
    \end{bmatrix}
    =
    \begin{bmatrix}
        0 & 1 \\
        -a & -b
    \end{bmatrix}
    \begin{bmatrix}
        \rho(t) \\
        \Delta_{\rho}(t)
    \end{bmatrix}
    ,
    \begin{bmatrix}
        \rho(0) \\
        \Delta_{\rho}(0)
    \end{bmatrix}
    \text{given,}
\end{equation}

\noindent
where the initial concentration, $\rho(0) = 0$, is referenced to the cytokine's basal level, and the initial rate of change, $\Delta_{\rho}(0)$, is stimulated by the TGN1412 infusion (see~\cite{Yiu2012}). Also, $a$ and $b$ are positive constants that express the sensitivity of the cytokine's acceleration to concentration and rate of change.

The cytokine's response modes are characterised by the eigenvalues, $\lambda_1$ and $\lambda_2$ (rad/day), of the stability matrix of the system, and this is as follows

\begin{equation}
\label{eq:cyt_storm}
    \begin{bmatrix}
        \dot{\rho}(t) \\
        \dot{\Delta_{\rho}}(t)
    \end{bmatrix}
    =
    \begin{bmatrix}
        0 & 1 \\
        -\lambda_{1}\lambda_{2} & (\lambda_{1} + \lambda_{2})
    \end{bmatrix}
    \begin{bmatrix}
        \rho(t) \\
        \Delta_{\rho}(t)
    \end{bmatrix}
    ,
    \begin{bmatrix}
        \rho(0) \\
        \Delta_{\rho}(0)
    \end{bmatrix}
    \text{given.}
\end{equation}



The parameters are chosen to minimize the error between the cytokine concentration and the clinical trial measurements performed in~\cite{Yiu2012}. For TNF-$\alpha$, $\lambda_{1} = \lambda_{2} = -2.63$ and $\Delta_{\rho}(0) = 32821$. In this work, we will be investigating the inflammatory effects of TNF-$\alpha$, as this cytokine is well-known for its pro-inflammatory properties.




\subsection*{Conduction Through a Myelinated Axon}
\label{subsec:myelin_model}

\begin{figure}[t]
    \centering
    \includegraphics[width=.5\columnwidth]{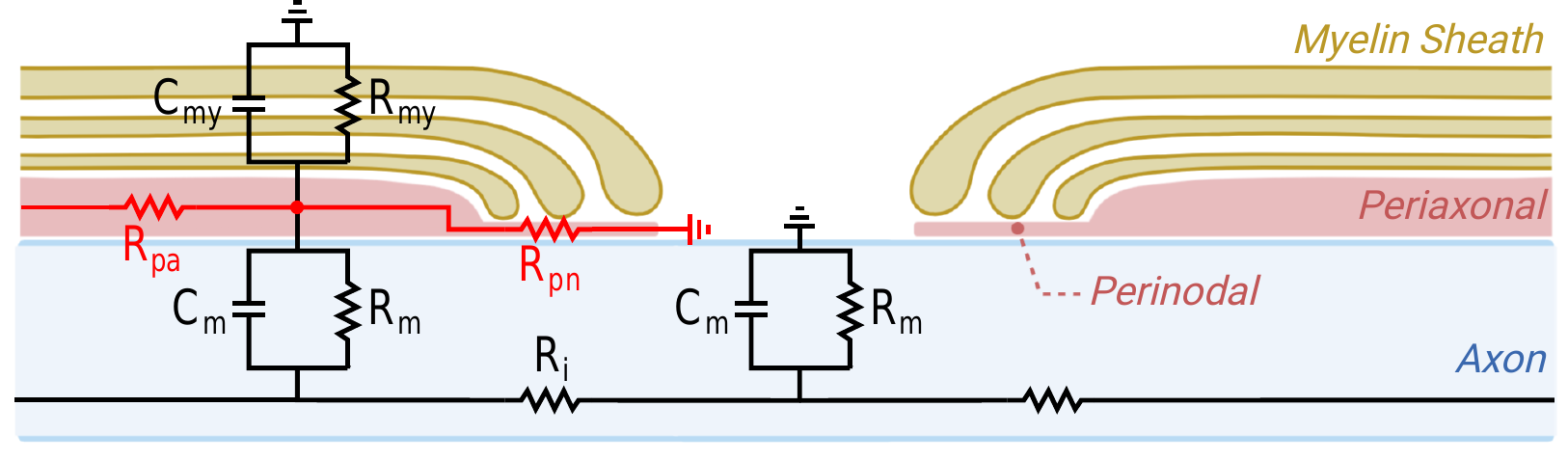}
    \caption{Schematic of the equivalent myelinated axon circuit, with added periaxonal ($P_a$) and paranodal ($P_n$) axial resistances~\cite{Cohen2020}.}
    \label{fig:long_section_axon}
\end{figure}

It is well known that some neurons contain a myelin sheath wrapped around sections of their axons. Myelin sheath helps propagate electrical impulses, known as action potentials (AP) or spikes, and avoid significant attenuation due to other synaptic processes. According to Cohen et al~\cite{Cohen2020}, the myelin sheath dynamics can be described using circuit theory. This circuit is then coupled with the Hodgkin-Huxley (HH) circuit model, which describes the AP dynamics in neurons (Figure~\ref{fig:long_section_axon}).

Axial resistance, $r_{i}$, of the axon core, can be defined as the ratio between axial resistivity, $R_{i}$, and the cross-sectional area, and this is represented as


\begin{equation}
    r_{i} = \frac{4R_{i}}{\pi d^{2}}.
\end{equation}

\noindent
where and $d$ is the axon core diameter. Let $\delta_{pa}$ be the periaxonal cross-sectional area, then

\begin{equation}
    \delta_{pa} = \frac{1}{2} \left[ -d + \sqrt{d^{2} + \left(\frac{4R_{pa}}{\pi r_{pa}}\right)}\right],
\end{equation}

\noindent
where the axial resistance in the periaxonal space, $r_{pa}$, is calculated as the ratio between periaxonal resistivity, $R_{pa}$, and periaxonal cross-sectional area. In this case, the axon core cylinder is surrounded {\color{black}by both the periaxonal and perinodal spaces forming a ``larger'' axon cylinder} of diameter $d + 2 \delta_{pa}$, thus

\begin{equation}
    r_{pa} = \frac{R_{pa}}{\pi \delta_{pa} (d + \delta_{pa})},
\end{equation}

\noindent
and the same calculation can be performed for $\delta_{pn}$, $R_{pn}$ and $r_{pn}$.

Recognising that a myelin sheath is in-series compaction of $n$ layers, the radial resistance of the sheath, $R_{my}$, is the sum of the resistance of each myelin membrane, $R_{mm}$ represented as

\begin{equation}
    R_{my} = \sum_{i=1}^{n}R_{mm_{i}},
\end{equation}
\noindent
and, the radial capacitance of the myelin sheath, $C_{my}$, may vary inversely to the sum of the capacitances of each of its composing membranes, $C_{mm}$, thus

\begin{equation}
    \frac{1}{C_{my}} = \sum_{i=1}^{n}\frac{1}{C_{mm_{i}}},
\end{equation}

\noindent
where the resistance and capacitance of a single myelin membrane are $R_{mm}$ and $C_{mm}$, respectively. Based on this the expression of the function can be represented in terms of the number of myelin lamellae, $n_{my}$, as follows

\begin{equation}
\label{eq:resist_nmy}
    n_{my} = \frac{R_{my}}{2 R_{mm}} = \frac{C_{mm}}{2 C_{my}}.
\end{equation}

For a more detailed analysis of the myelin sheath's modelling, the reader is referred to the work of Cohen et al~\cite{Cohen2020}.





\subsection*{Cytokine-induced Demyelination}
\label{subsec:cyt_demy}

As mentioned in the Introduction, a cytokine storm is released to fight infections and, in the process, can damage healthy tissues. There are numerous pieces of evidence linking cytokine storms to neurodegeneration~\cite{Waito2016, Yiu2012, Ludwig1268, Bitsch2000}. However, to the best of our knowledge, none of them goes as far as linking the storm's intensity to an approximate number of myelin sheaths. {\color{black}Applying linear regression ($R=0.508$, $p=0.001$)} on the data provided by Ludwig et al~\cite{Ludwig1268}, we establish a proportional relationship between the severity of the neuropathy ($\zeta$) and the serum concentration of TNF-$\alpha$, $\rho(t)$, as

\begin{equation}
\label{eq:cyt_demy}
    \zeta = \frac{\rho(t) - 8.374}{1.761},
\end{equation}


\noindent
and in this relationship, the stronger the cytokine storm is, the more severe the degeneration. The severity of the neuropathy was defined according to \cite{Ludwig1268} as a scoring function, where scoring nerve functions of `$0$' is for typical values of the nerve, `$1$' for affected nerves (either decreased amplitude and/or decreased nerve conduction velocity) and `$2$' for no stimulation possible. The severity that initially ranges between $0$ and $8$ in total should be re-scaled to match a normal $n_{my}$ as reference.

\subsection*{A Linear Model of Demyelination}
\label{subsec:lin_model}

Based on a given healthy neuron, we want to predict the signal effects on spike trains that are expected from a demyelinated neuron. Let $n$, the number of the myelin sheaths, be a tunable parameter of the model (in this section, we refer to $n_{my}$ as $n$ for clarity of index notation). This mechanism is depicted in the block diagram in Figure~\ref{fig:viral_persistence_demyelination}(b). 

Observing faulty signals resulting from the propagation of standard signals through a transfer function superimposed on the original system is a well-established concept; examples of its application include modelling of cables \cite{pintelon1990identification}. It makes intuitive sense to observe such a transfer function as taking the health system's output and delivering the faulty, unhealthy signal as the faulty output, which results in signal degradation for a worsening channel it traverses. The opposite process, in which an ``adaptor" transfer function would convert a signal from a deteriorated neuron into one of a healthy neuron, is anti-causal as it would have to introduce negative time delays in the signal.

\begin{figure}[t]
    \centering
    \includegraphics[width=.5\columnwidth]{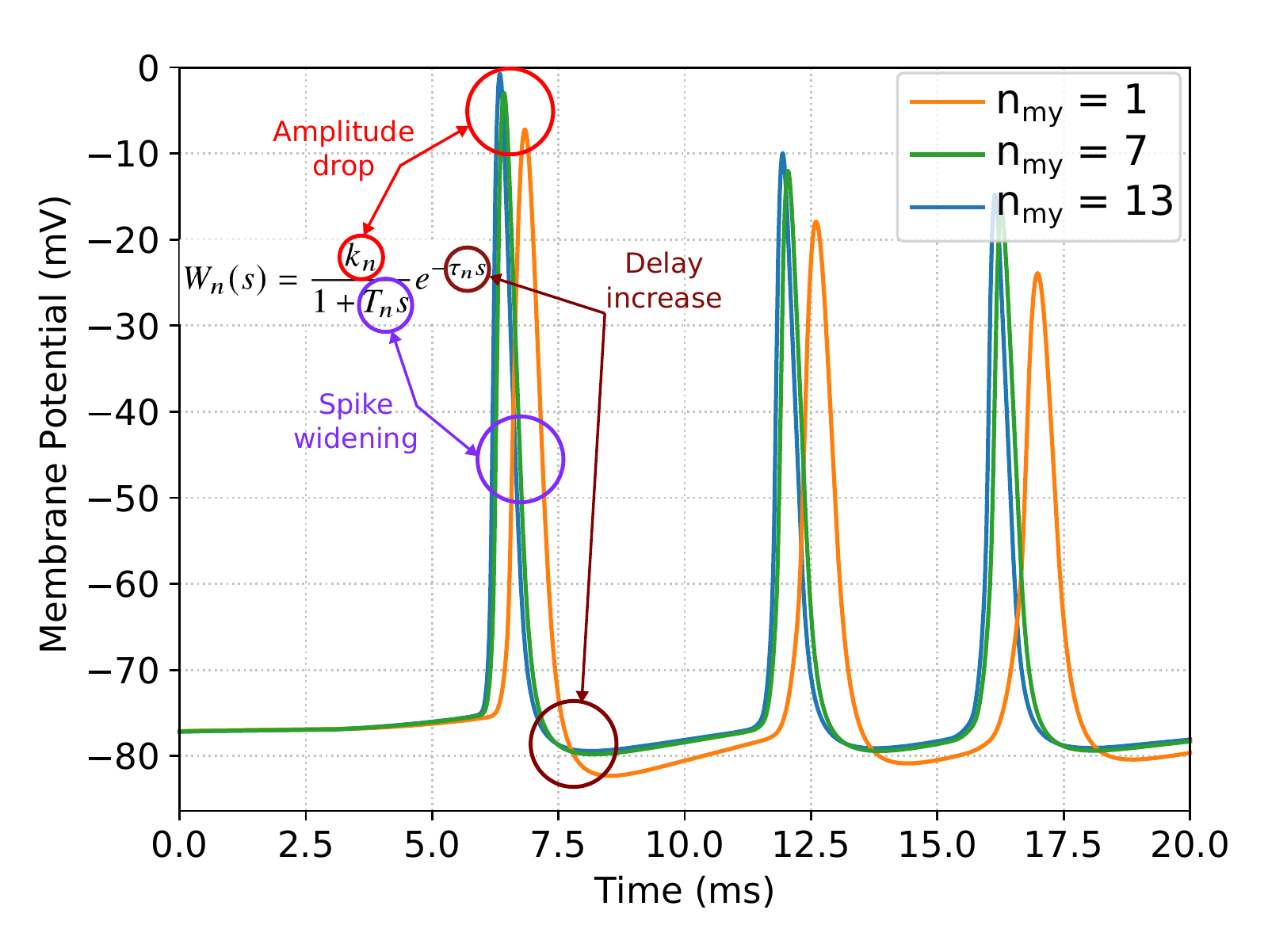}
    \caption{Mapping of the model components and the effects on the propagated signal, and illustration of signal behaviour as it transitions from healthy to unhealthy state.}
    \label{fig:logic}
\end{figure}

Now that we have established the processing chain, the choice for the transfer function is made by observing the general trends in the output signals for various values of $n$. Namely, as seen in Figure~\ref{fig:logic} the transfer function emulating the effect of myelin deficiency needs to allow for the attenuation of the signal, widening of the spikes, and overall propagation delay. An obvious candidate is the traditional First Order Plus Time Delay (FOPTD) function, given by

\begin{equation}
\label{eq:tf_FOPTD}
    W_n(s)=\frac{k_n}{1+T_ns}e^{-\tau_n s},
\end{equation}

\noindent
where we assume that the parameters $k_n$, $T_n$ and $\tau_n$ have different values for different values of $n$. While those values could be tabulated and looked up for specific values of $n$, we set a more ambitious goal of determining the laws according to which they change. Here we make a hypothesis that they follow exponential law (namely, $\log k_n=a_0\cdot a_r^n$, $T_n=T_o\cdot T_r^n$ and $\tau_n=\tau_0\cdot\tau_r^n$), based on the following reasoning.

Turning a healthy signal into a deteriorated one in our model is a process of cancelling the effect of passing through some of the sheathes. For example, if the healthy baseline signal is achieved with $n=13$, emulation of $n=10$ can be interpreted as undoing the effect of $\Delta n=3$ sheathes, i.e. the effect of a cascade of three blocks representing a "sheath undo function". This approximation is good for larger $\Delta n$, as it allows for the approximations like $(1+T\bar{s})^{\Delta n}\approx 1+T^{\Delta n} s$, i.e. giving rise to a $T_n\sim\exp(n)$ law. In the Results and Discussion section, we revisit this hypothesis, both in terms of the exponential law's existence and in terms of the domain of accuracy. Introducing exponential laws for the coefficients in this transfer function gives us the final form of our transfer function, which is represented as

\begin{equation}
\label{eq:tf_exp}
W_n(s)=\frac{e^{a_0\cdot a_r^n}}{1+T_0\cdot T_r^ns}e^{-\tau_0\cdot \tau_r^n s}=\frac{e^{a_0\cdot a_r^n-\tau_0\cdot \tau_r^n s}}{1+T_0\cdot T_r^ns}.
\end{equation}

The final form of the equation \ref{eq:tf_exp} suggests the rationale behind suggesting exponential behaviour of $a=\log k$ instead of $k$ itself: in the $s$-domain, we obtain a function of the form $\frac{\exp(\alpha+\beta s)}{\gamma+\delta s}$, which in turn in Fourier domain corresponds to $\frac{\exp(\alpha+j\beta\omega)}{\gamma+j\delta\omega}$. Knowing that the behaviour of the myelin circuit originates from connections of $R$ (purely passive, real impedance) and $C$ (purely active, imaginary impedance), it is expected to observe this symmetric real-imaginary coupling of terms.

FOPTD is not an uncommon choice in biophysics, where it has been used to model glucose control~\cite{farmer2009effectiveness}, because they are simple for identification~\cite{sundaresan1978estimation} and for quick and accurate tuning of the controllers that can regulate their behaviour~\cite{sree2004simple}. Nonetheless, our model may help in designing chemical control loops for myelin reinforcement in a similar manner.

To verify the quality of the model, we introduce a metric based on root mean square error (RMSE):

\begin{equation}
\label{eq:metrice}
    M_n= 20\log_{10}\left(\frac{\text{\textit{RMSE}}_{N,n}}{\text{\textit{RMSE}}_{W,n}}\right).
\end{equation}

Here, $\text{\textit{RMSE}}_{W,n}$ stands for the RMSE of the output of our transfer function $W_n$ compared to the actual output signal for $n$ sheaths, i.e. if the $m$ is the number of samples, which is represented as

\begin{equation}
    \text{\textit{RMSE}}_{W,n} = \sqrt{\frac{1}{m}\sum_{i\ge1}(x_{n,i} - x_{W,i})^2}.
\end{equation}
\noindent
Analogously, we can find the value for $\text{\textit{RMSE}}_{N,n}$ which corresponds to the RMSE of the output signal produced by another model $N$ compared to the $n$th actual output signal. The quantity $M_n$ is positive where our model $W_n$ is more accurate (has lower RMSE) than the model $N$ we are comparing to it.

\subsection*{Signal Analysis}

Visually, we first noticed subtle shifts in amplitude (peak potential reached by the membrane) and in time (spikes were taking longer to reach their peak values). We then decided to quantify these shifts, both in amplitude and in time, on average, by proposing a metric that we called \textbf{relative mean shift}. For the analysis on time shift, we consider the points in time where each spike peaked at the input as $T_{in}^{k}$, where $k = \{1, 2, 3, ..., n\}$ identifies the order of each spike and, $T_{out}^{k}$ as the peak times at the output. Thus, we define the relative mean time shift, $\overline{\delta_{t}}$, as

\begin{equation}
\label{eq:time_shift}
    \overline{\delta_{t}} = \frac{1}{n}\sum_{k = 1}^{n} (T_{in}^{k} - T_{out}^{k}),
\end{equation}

\noindent
analogously, we can define the relative mean amplitude shift, $\overline{\delta_{v}}$, with $V_{in}^{k}$ and $V_{out}^{k}$ as the peak amplitudes of spike $k$ at the input and output, respectively.

As there is an average shift in time inside the channel itself due to demyelination, we also expect an increase in \textbf{latency} as we decrease $n_{my}$. In order words, latency is the time interval between the input and the output, and it often occurs due to the channel or network's intrinsic characteristics. In our scenario, we are looking for the time the first spike peaked at the output, concerning the point when this same spike peaked in the soma of the neuron. Later, analogous to the relationship between $\overline{\delta_{t}}$ and the latency, we decided to look into a potential attenuation on the power of the signal, $P$, as we have already discussed a relatively heavy mean attenuation in the membrane potential, $\overline{\delta_{v}}$. Let us define $P$ as

\begin{equation}
\label{eq:sig_power}
    P = \displaystyle\lim_{N \to \infty} \left( \frac{1}{2N + 1} \sum_{n = -N}^{N} |x[n]|^{2} \right),
\end{equation}

\noindent
where in a set of $N$ samples, $x[n]$ corresponds to the potential of the membrane at the $n$-th sample.

As there are changes in specific time points where spikes peak, it is most likely that there will be changes in the spiking rate. For that reason, we calculated the approximated spiking rate from the mean inter-spike interval (ISI). Consider a spike train with $N$ spikes and, between each spike there is an interval of value $t_{n}$ where $n = \{1, 2, 3, ..., N-1\}$. We can define an equation for calculating the spiking rate ($\lambda_{s}$) from the ISI, thus

\begin{equation}
\label{eq:spiking_rate}
    \lambda_{s} = \frac{1}{N-1}\sum_{n = 1}^{N-1} t_{n},
\end{equation}

\noindent
which, according to Yu et al~\cite{Yu2012}, there is a linear relationship ($R^{2} = 0.9304$) between the spiking rate, $\lambda_{s}$, and the probability of releasing glutamate into the synaptic cleft, $P(r)$. Glutamate is an excitatory neurotransmitter that is present in most synaptic connections. Its deficit can dysregulate the depolarisation of the postsynaptic neuron and, as a consequence, compromise the transfer of information in a neuronal network. This relationship is expressed as

\begin{equation}
\label{eq:glut_prob}
    P(r) = 0.038 \cdot \lambda_{s} + 0.14.
\end{equation}

We also quantified the attenuation of the signal, which is the gradual loss of power of a signal over its propagation through the channel. Depending on the attenuation coefficient, one can calculate more accurately the attenuation in a specific material. For our analysis, we used the generic form of attenuation for RF cables. This decision is based on the fact that the axonal pathway is modelled using cable theory as a leaky cable. Thus,

\begin{equation}
\label{eq:attenuation}
    A[\text{dB/100m}] = 10 \cdot \log_{10}\left(\frac{P_{i}[W]}{P_{o}[W]}\right),
\end{equation}

\noindent
where $P_{i}$ and $P_{o}$ are the input and output power of the signal. In this analysis, the input is the spike train through a healthy neuron and the output would be the spike train through a demyelinated one.

Lastly, we also analysed the relation between the reference spike train ($n_{my} = 13$) and all other demyelinating scenarios. The intention is to understand how much power is being transferred between each pair of signals. With that in mind, we applied a coherence ($C_{xy}$) metric, which is described as

\begin{equation}
\label{eq:sig_coherence}
    C_{xy}(\omega) = \frac{S_{xy}(\omega)^{2}}{S_{xx}(\omega)S_{yy}(\omega)},
\end{equation}

\noindent
where $S_{xy}(\omega)$ is the cross-spectral density between the two signals and, $S_{xx}(\omega)$ and $S_{yy}(\omega)$ are the power spectrum densities of input and output, respectively.

\section*{Results and Discussion}
\label{sec:res_disc}

\begin{figure}[t]
    \centering
    \subfigure[][Relative mean time and amplitude shifts.]{%
    \label{fig:mean_shift}%
    \includegraphics[width=0.325\textwidth]{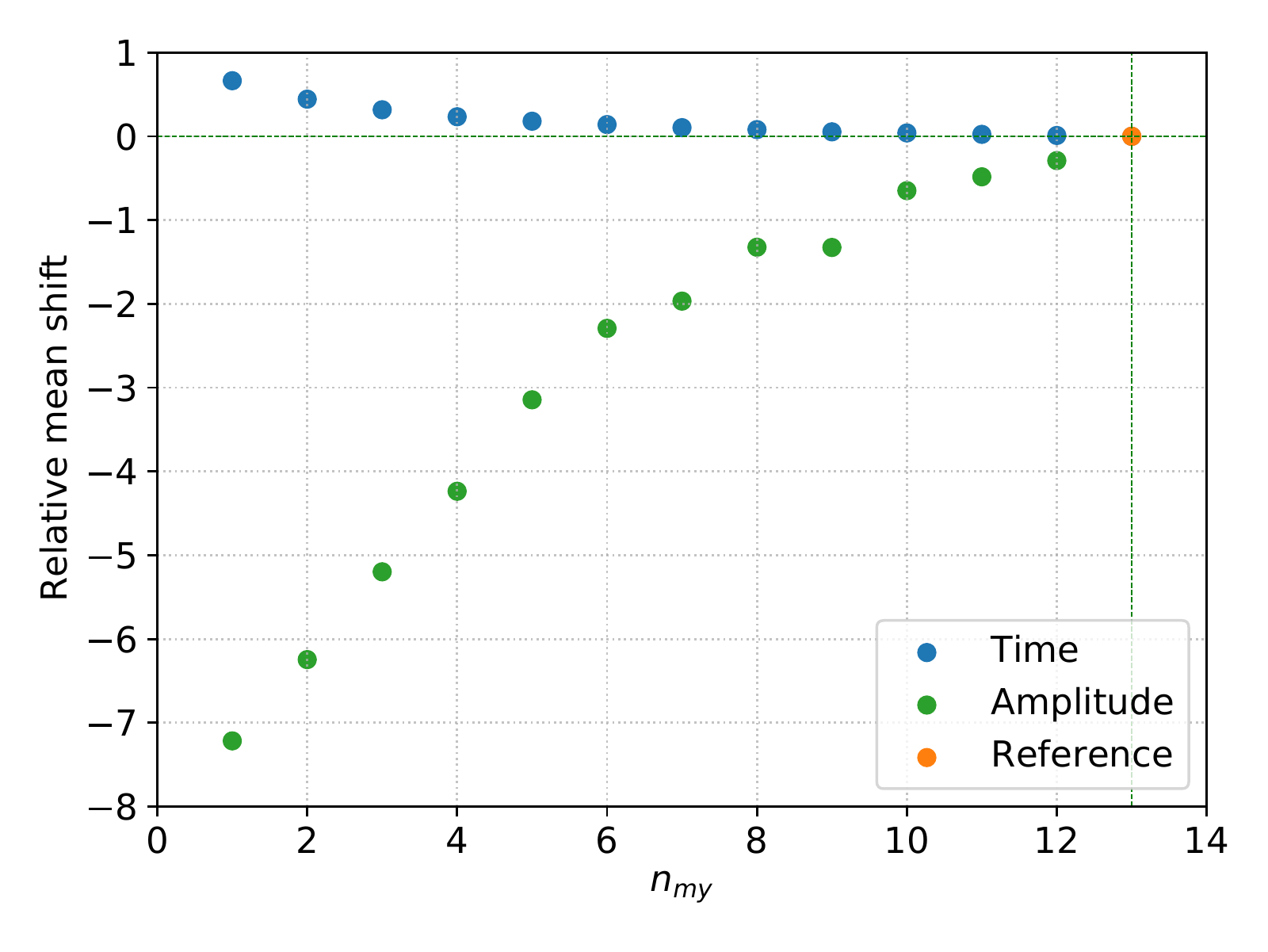}}%
    ~
    \subfigure[][Relative mean latency.]{%
    \label{fig:mean_latency}%
    \includegraphics[width=0.325\textwidth]{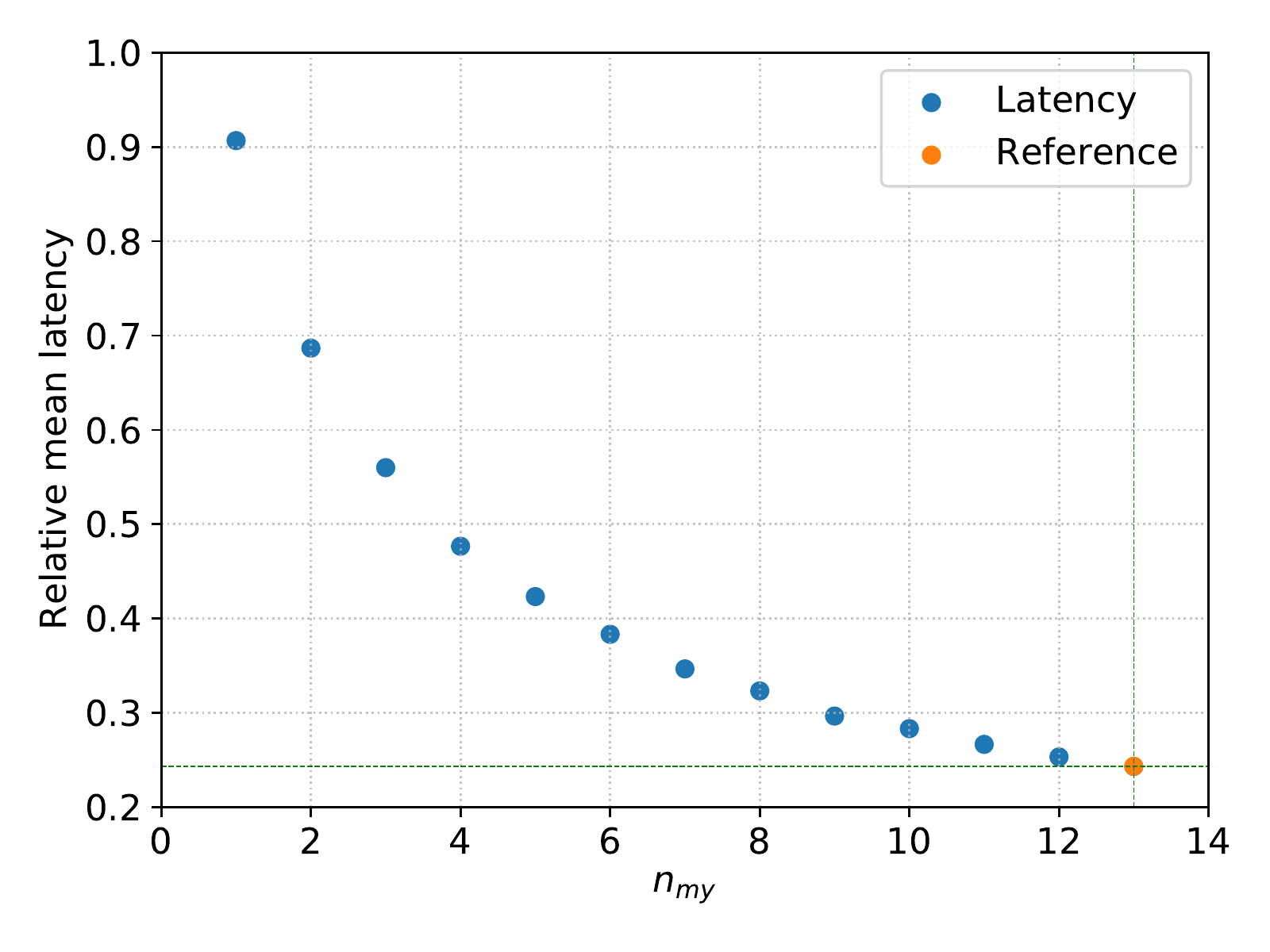}}%
    ~
    \subfigure[][Signal Power.]{%
    \label{fig:signal_power}%
    \includegraphics[width=0.325\textwidth]{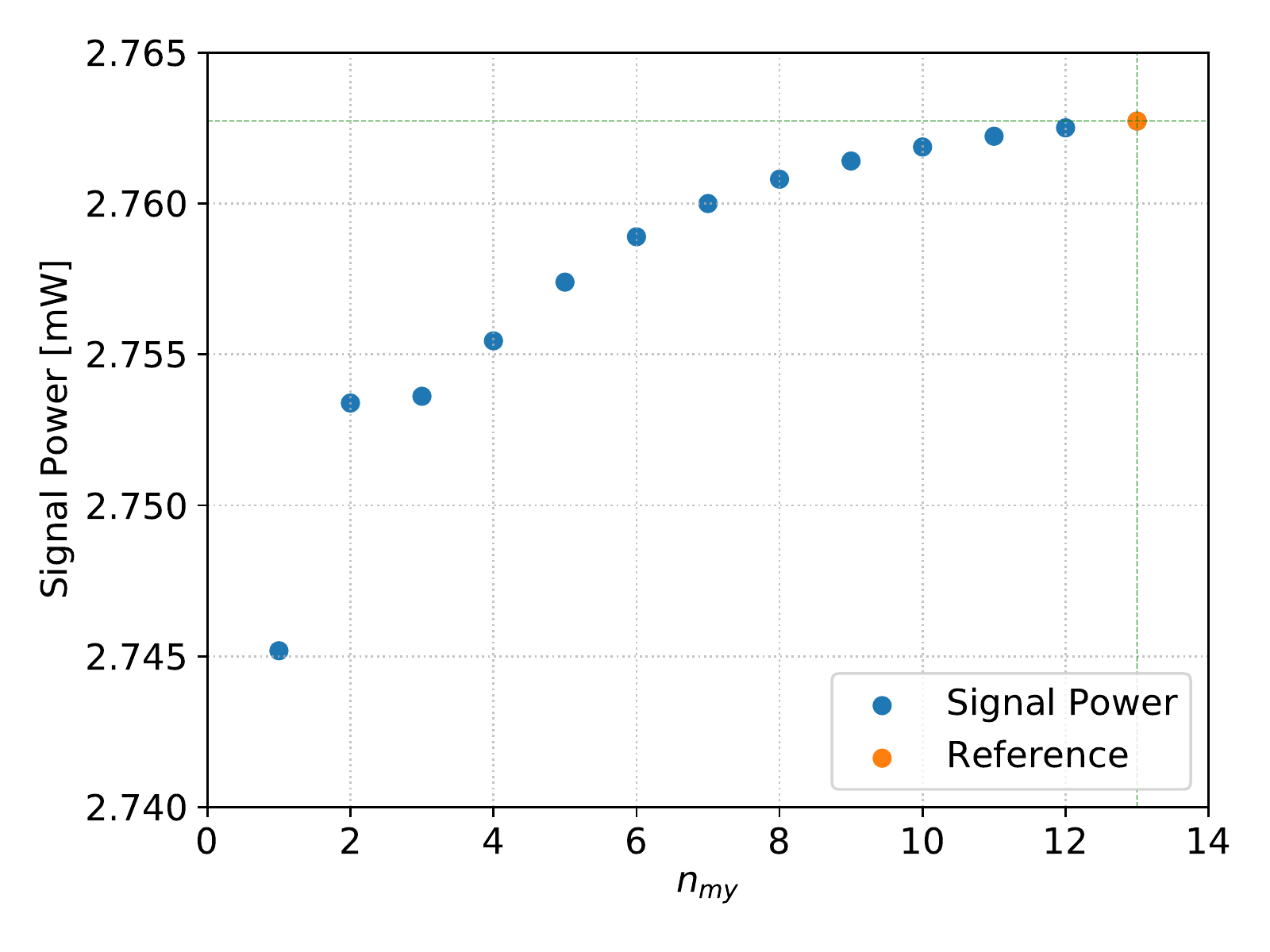}}%
    \\
    \subfigure[][Spiking rate.]{%
    \label{fig:spiking_rate}%
    \includegraphics[width=0.325\textwidth]{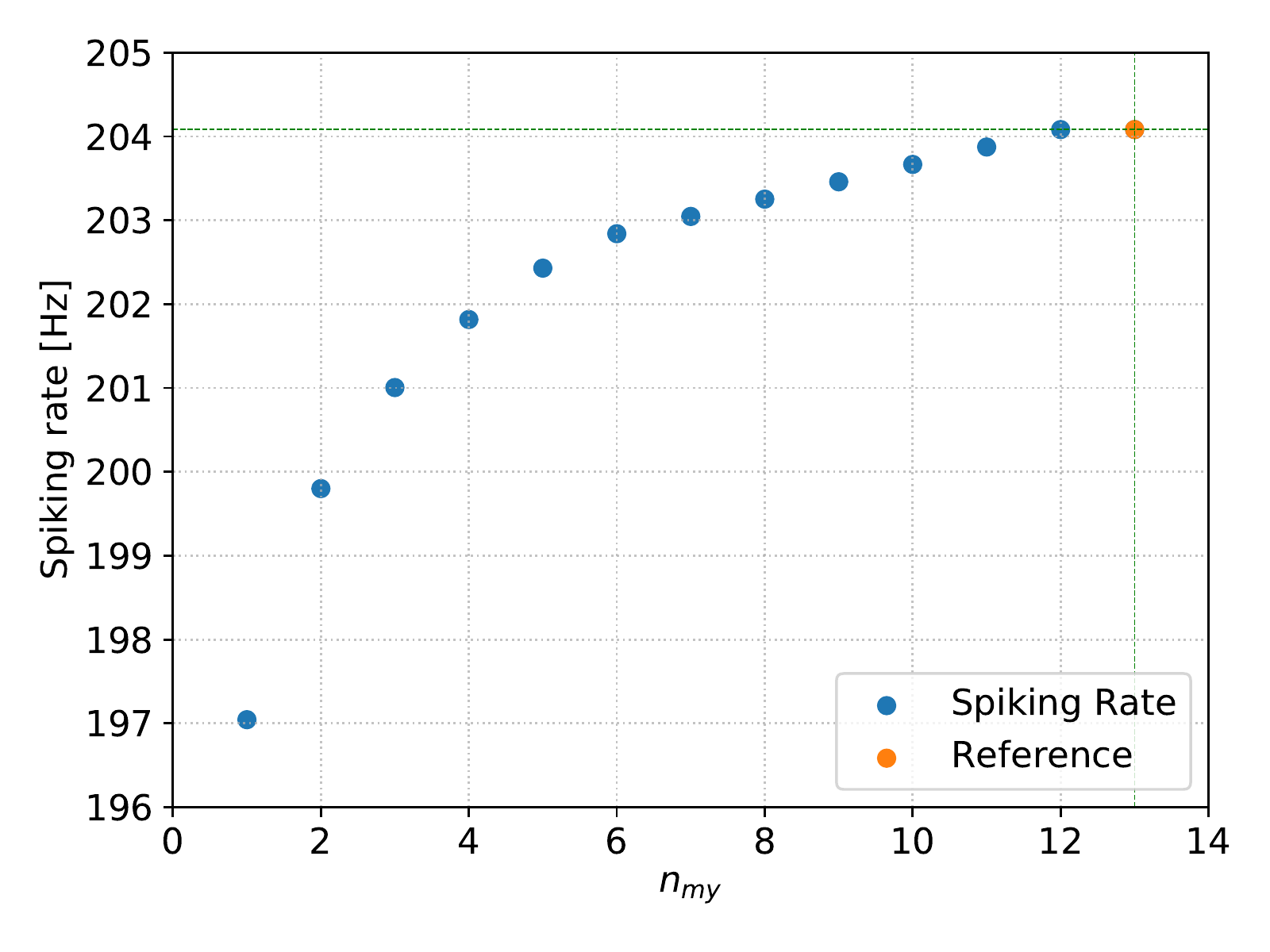}}%
    ~
    \subfigure[][Glutamate release probability.]{%
    \label{fig:glutamate_release_probability}%
    \includegraphics[width=0.325\textwidth]{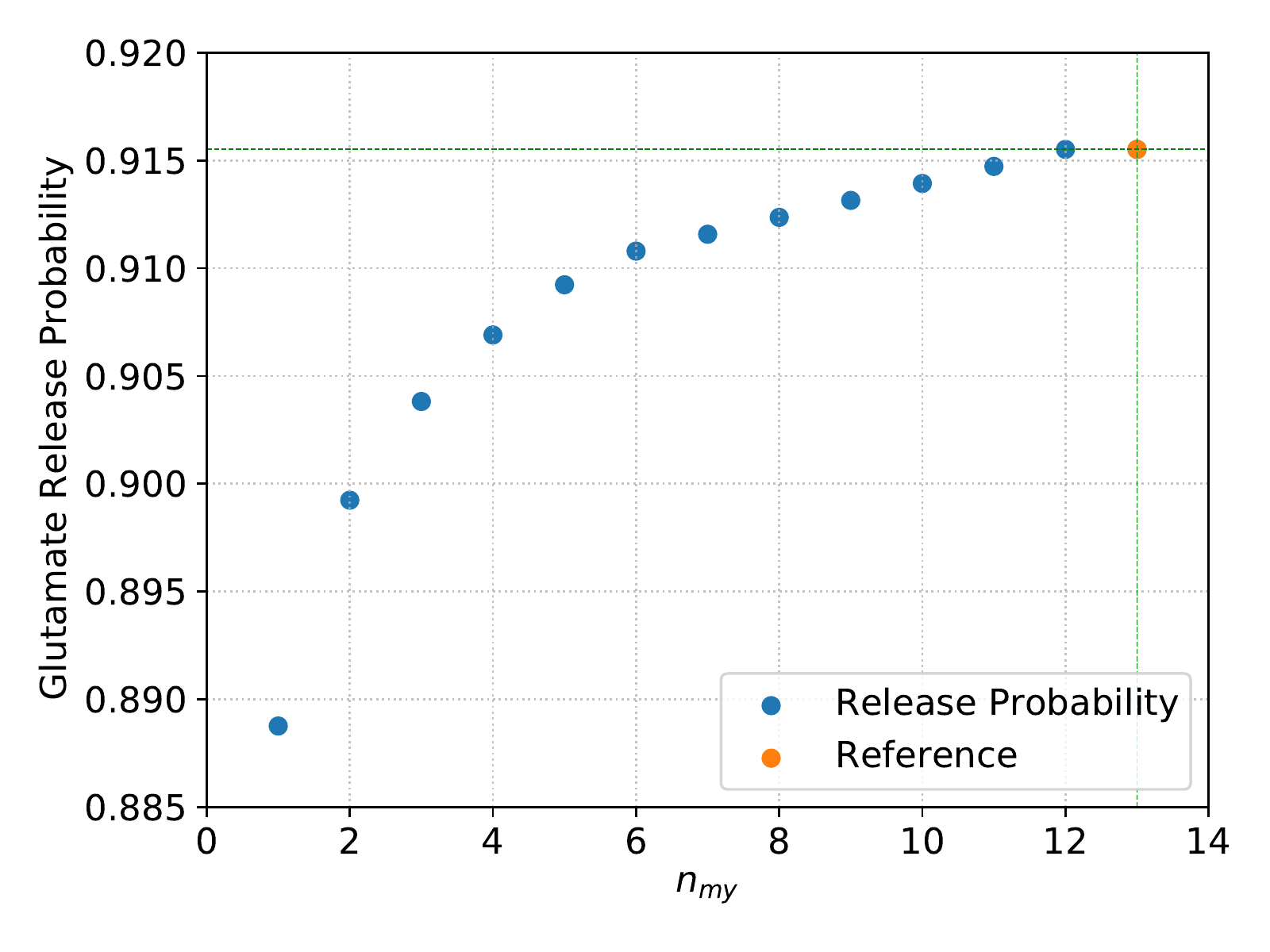}}%
    ~
    \subfigure[][Attenuation.]{%
    \label{fig:attenuation}%
    \includegraphics[width=0.325\textwidth]{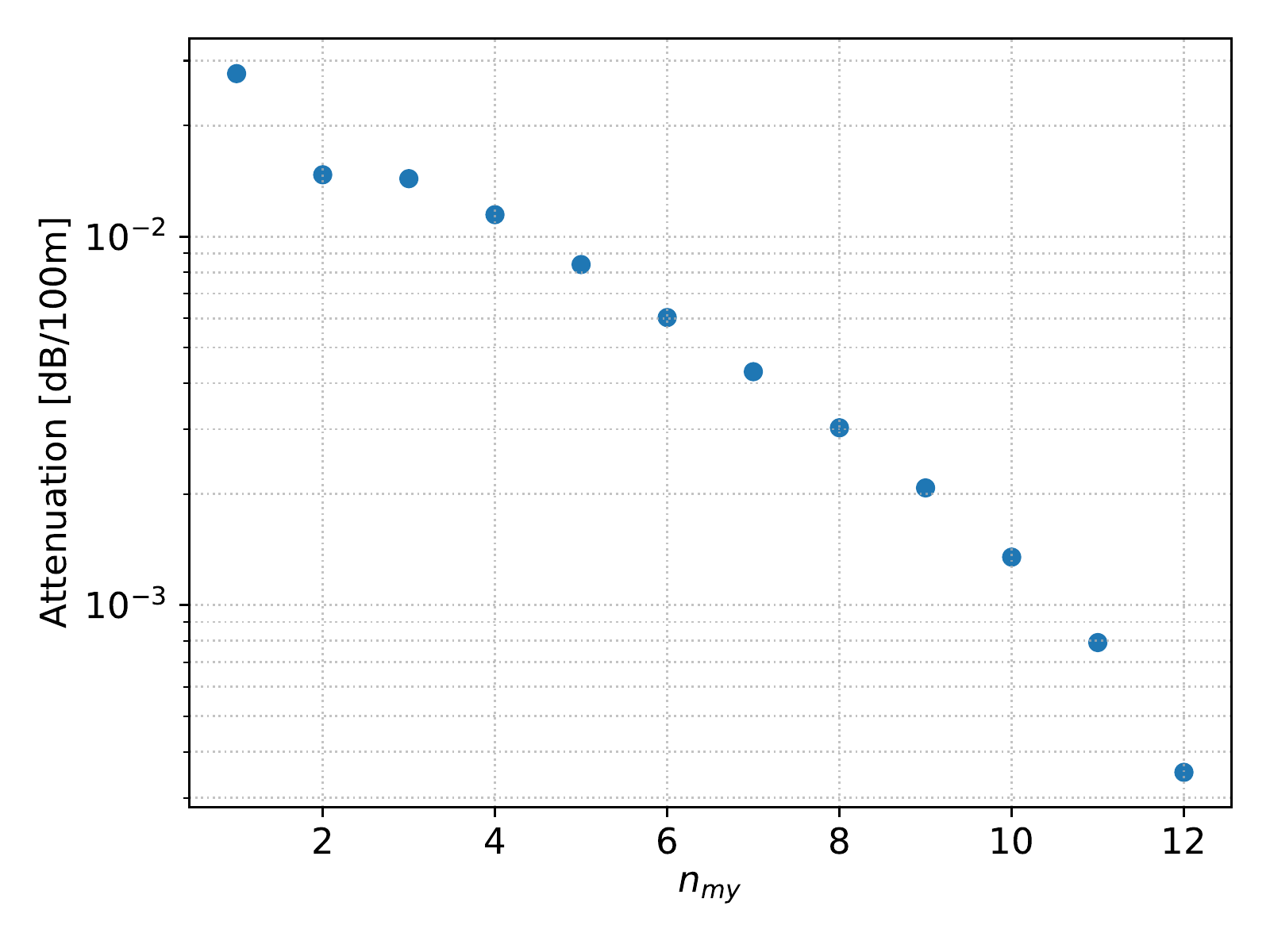}}%
    
    \caption{Measurements of attenuation and delay. All results except for Figure~\ref{fig:mean_latency}, use the signal $n_{my} = 13$ as their input and $1 \le n_{my} \le 12$ as the output.  Figure~\ref{fig:mean_latency} uses the recordings from the soma of the neuron and is compared with $1 \le n_{my} \le 13$.}
    \label{fig:n_my_influence}
\end{figure}

Let us consider the proportionality between the severity of the neuropathy and the number of myelin lamellae, $n_{my}$, as described in the Cytokine-induced Demyelination section. Our analysis consisted of evaluating the neuronal behaviour and spike propagation under normal circumstances ($n_{my} = 13$), followed by an analysis on the demyelination by decreasing $n_{my}$. The objective is to understand what happens to the neuronal information when travelling through a demyelinated axonal pathway. The neuron receives an external current of $3$ nA for $15$ ms at time $t = 2.5$ ms in a $20$-ms simulation. The spikes evoked under normal circumstances are considered our input of the channel, and the spikes at the far end of the axon are the output. The axon itself is our communication channel, while the demyelination is acting as an attenuator for the channel (Figure~\ref{fig:viral_persistence_demyelination}(b)).

\subsection*{Analysis of a Demyelination-induced Channel Attenuation}

The results for $\overline{\delta_{t}}$ and $\overline{\delta_{v}}$ from Equation~(\ref{eq:time_shift}) are shown in Figure~\ref{fig:mean_shift}. It shows that, although \textit{time} has quite slight delays as we decrease $n_{my}$, the \textit{amplitude} of the membrane potential is a bit more drastically attenuated. In other words, $\overline{\delta_{v}}$ is almost \textbf{eight times} more degraded than $\overline{\delta_{t}}$ in the worst scenario ($n_{my} = 1$). This matches fundamental computational neuroscience theory on neuronal modelling which states that a neuron gradually leaks a small amount of the input signal as it travels through it, and this ``leak'' worsens on unmyelinated cables~\cite{eliasmith2003neural, Hamada2017}. Each curve's shape resembles a capacitor charging (amplitude) or discharging (time) in RC circuits. This comes as no surprise since both the axonal compartment and the myelin sheath are roughly expressed as RC circuits (Figure~\ref{fig:long_section_axon}). Furthermore, as latency is an operation on time of the peak, it is expected the curve behaves similarly as it does for $\overline{\delta_{t}}$, as shown in Figure~\ref{fig:mean_latency}. This time, we notice that there is a subtle ``lag'' for the signal to travel across the axon even under normal circumstances {\color{black}matching findings in the literature~\cite{Bando2008} for models of CNS demyelination}. This is most likely due to a maximum conduction speed inherent in the axonal membrane itself. As we go from our worst scenario towards a regular healthy myelin sheath, there is a massive decrease of about $73\%$ in the latency.

 On the other hand, Figure~\ref{fig:signal_power} shows that the decrease in signal power is quite subtle, and from the worst-case scenario to the best, there is a difference of less than $20\,\mu$W. This indicates how demyelination affects the energy consumption per unit time used to propagate the axon's action potentials. As the signal starts to get degraded, it is less and less likely a spike would be evoked at the post-synaptic neurons connected to a demyelinated cell~\cite{Hamada2017}. Demyelination does not affect only the post-synaptic neuron by reducing the chances of evoking post-synaptic potential, but it can compromise the spiking rate of the demyelinated neuron itself. Figure~\ref{fig:spiking_rate} depicts how the spiking rate is affected by the demyelination. As we expected, as the spikes start to get wider and far from each other, the spiking rate gets lower. {\color{black}This corroborates findings on demyelination-induced effects on spiking rate~\cite{Coggan2010}.} The rate at which a neuron fires action potentials is significant for modulating and encoding neuronal information in cognitive, sensory and motor functions.

Not surprisingly, when we calculate the probability for releasing glutamate in Figure~\ref{fig:glutamate_release_probability}, it shows that the probability curve has the same shape as the spiking rate when increasing $n_{my}$. This is due to the linear relationship expressed in Equation~(\ref{eq:glut_prob}). There is a subtle decrease in $P(r)$, which may be aggravated depending on how a specific neuron encodes the information. {\color{black}This matches findings in the literature that state axonal coding is very sensitive to subtle changes in the pattern of the stimuli~\cite{Shrager1993}.}

Furthermore, we decided to investigate the attenuation caused by the demyelination from a more generic communication systems point of view as expressed in Equation~\ref{eq:attenuation}. The results presented in Figure~\ref{fig:attenuation} show how the signal is more attenuated as we remove each myelin sheath. {\color{black}It not only shows consistency between our results and validates our hypothesis but also supports findings on failing pre-synaptic AP due to demyelinating diseases~\cite{Hamada2017}.}

\begin{figure}[t]
    \centering
    \subfigure[][Signal coherence with $n_{my} = 1$.]{%
    \label{fig:coh-a}%
    \includegraphics[width=0.325\textwidth]{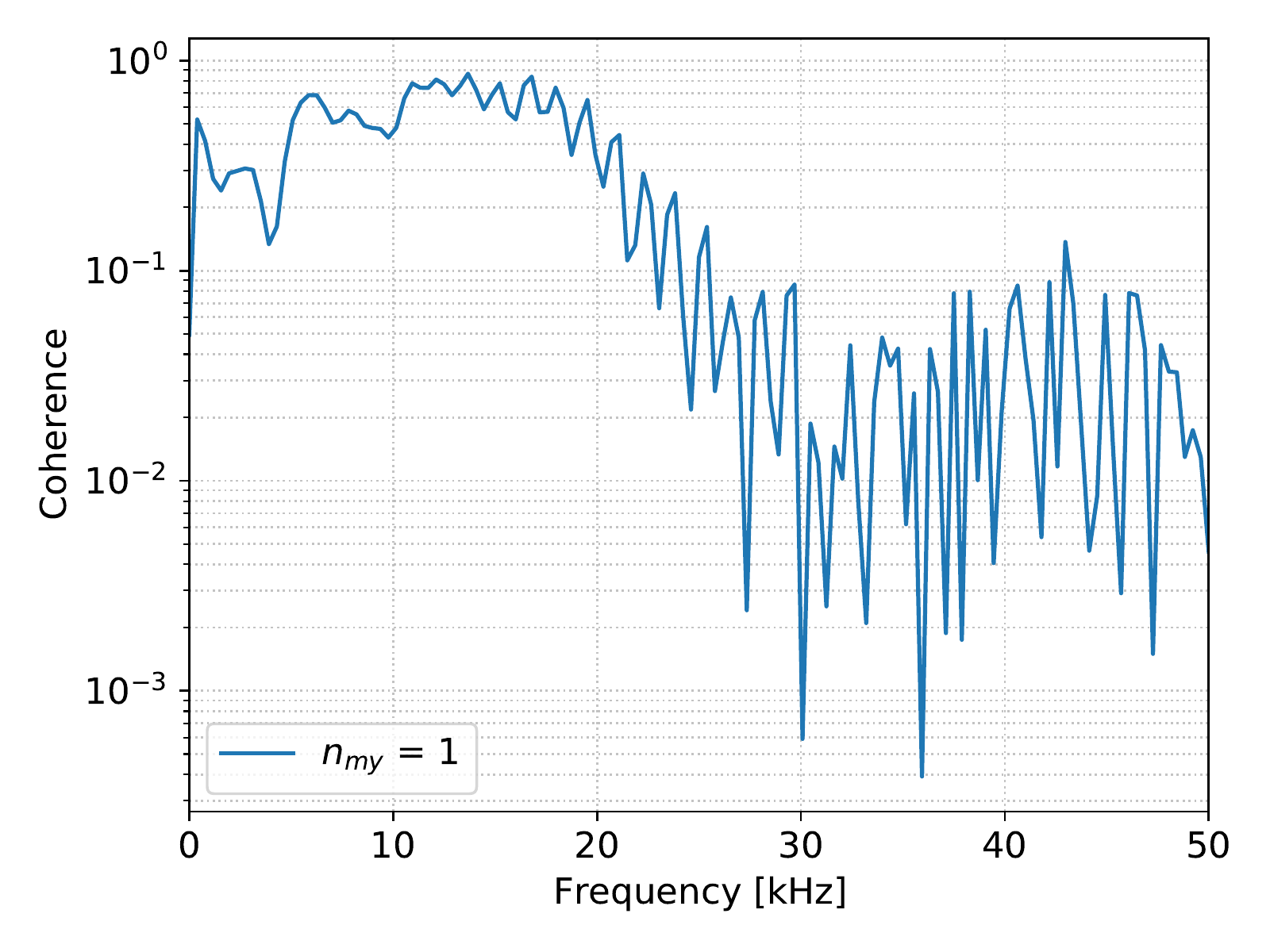}}%
    ~
    \subfigure[][Signal coherence with $n_{my} = 12$.]{%
    \label{fig:coh-b}%
    \includegraphics[width=0.325\textwidth]{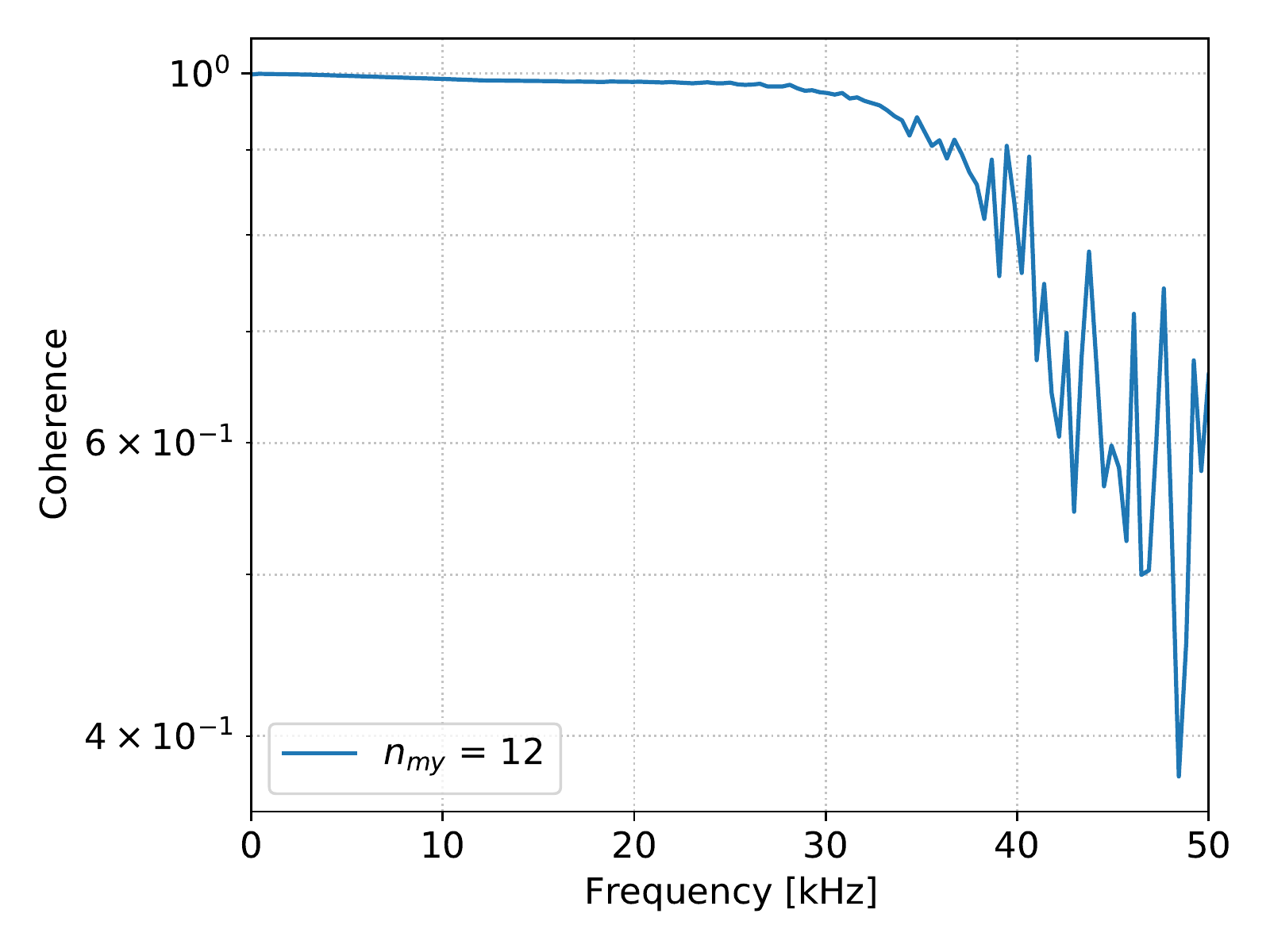}}%
    ~
    \subfigure[][Mean and standard deviation of the signal coherence for all $n_{my}$.]{%
    \label{fig:coh-c}%
    \includegraphics[width=0.325\textwidth]{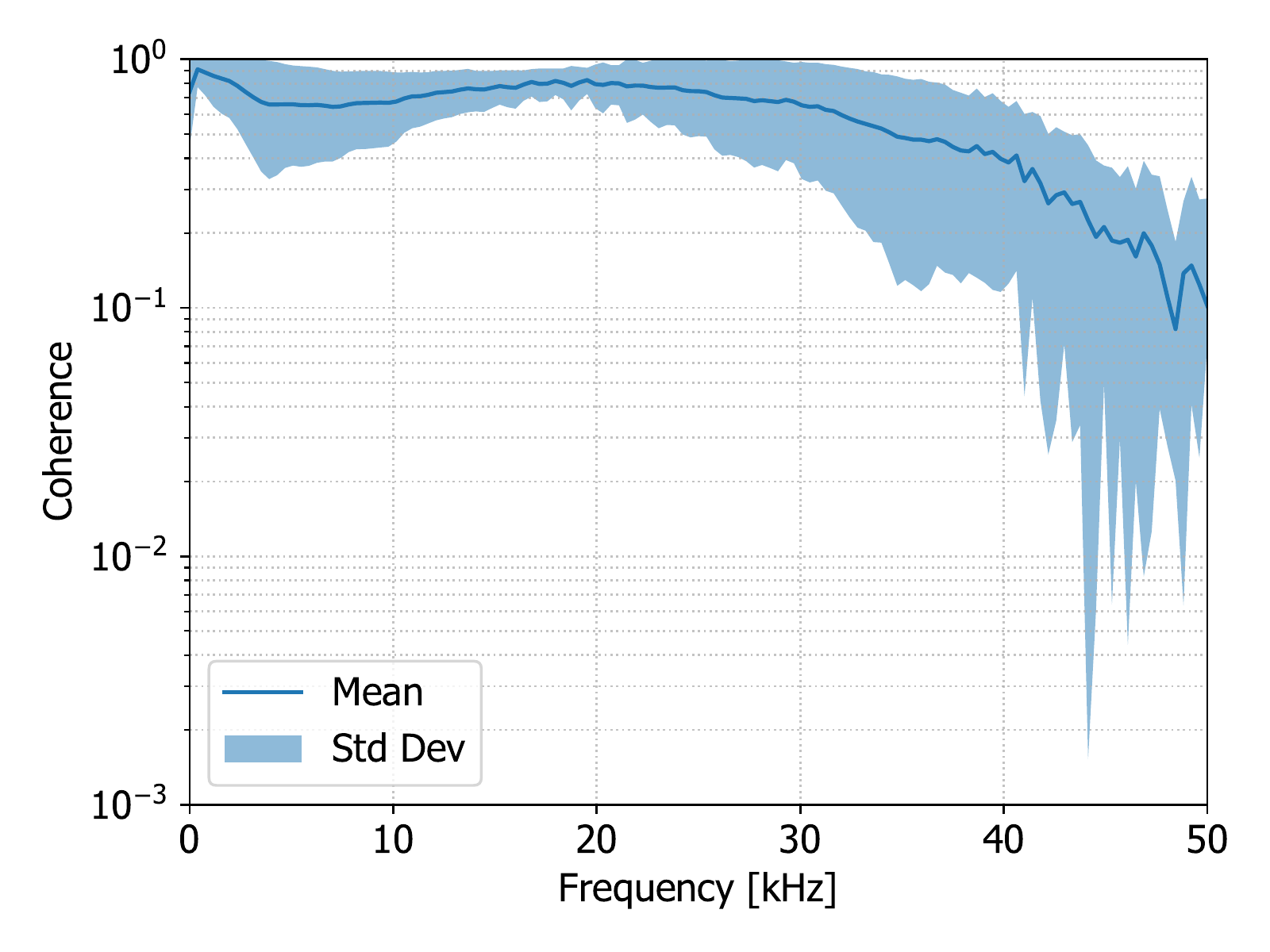}}%
    \caption{Signal coherence between $n_{my} = 13$ and (a) $n_{my} = 1$; (b) $n_{my} = 12$; and, (c) mean and standard deviation of the signal coherence with $1 \le n_{my} \le 12$.}
    \label{fig:signal_coherence}
\end{figure}

Lastly, Figure~\ref{fig:signal_coherence} shows the results for the signal coherence. {\color{black}In Figure~\ref{fig:coh-a}, we show the coherence with regards to our worst scenario, $n_{my} = 1$, in which there clearly are more oscillations in lower frequency bands when compared to Figure~\ref{fig:coh-b}. Figure~\ref{fig:coh-b} shows the coherence for a light demyelination, $n_{my} = 12$, where only $1$ sheath has been removed with regards to a healthy scenario, $n_{my} = 13$. Figure~\ref{fig:coh-c}, describes the mean and standard deviation of how the coherence values ($1 \leq n_{my} \leq 12$) fluctuate in the $0-50$kHz spectrum. Neuronal coherence has been known to serve neuronal communication as an indicator of the efficiency of the exchange of information~\cite{Schoffelen111}. All coherence plots showed some fluctuations for the higher end of the frequency range.} We believe the fluctuations in the coherence plots is a finite window effect. In other words, the Fast Fourier Transform (FFT) implicitly filters the data with a rectangular time-domain filter, which is a sinc-shaped filter in the frequency domain. For this reason, we are bound to get those lobes.

\subsection*{The Linear Model Identification and Verification}

Let us observe the changes that the output of a neuron goes through when $n$ varies, and understand these dynamics as depicted in Figure~\ref{fig:logic}. Reduction in the number of myelin sheaths ($n$, $1 \le n\le 13$) causes an increasing delay in the signal, i.e., spikes start later in neurons with less myelin, and they take longer to reach the peak value. On the other hand, in terms of the shape of the spikes, we observe the effect on the spike height and the spike width (full width at half maximum, FWHM) also in Figure~\ref{fig:logic}. This solution was adopted by identifying a suitable transfer function.

\begin{figure}[t]
\centering
    \subfigure[][Time offset of spike peaks]{%
    \label{fig:tosp}%
    \includegraphics[width=0.4\columnwidth]{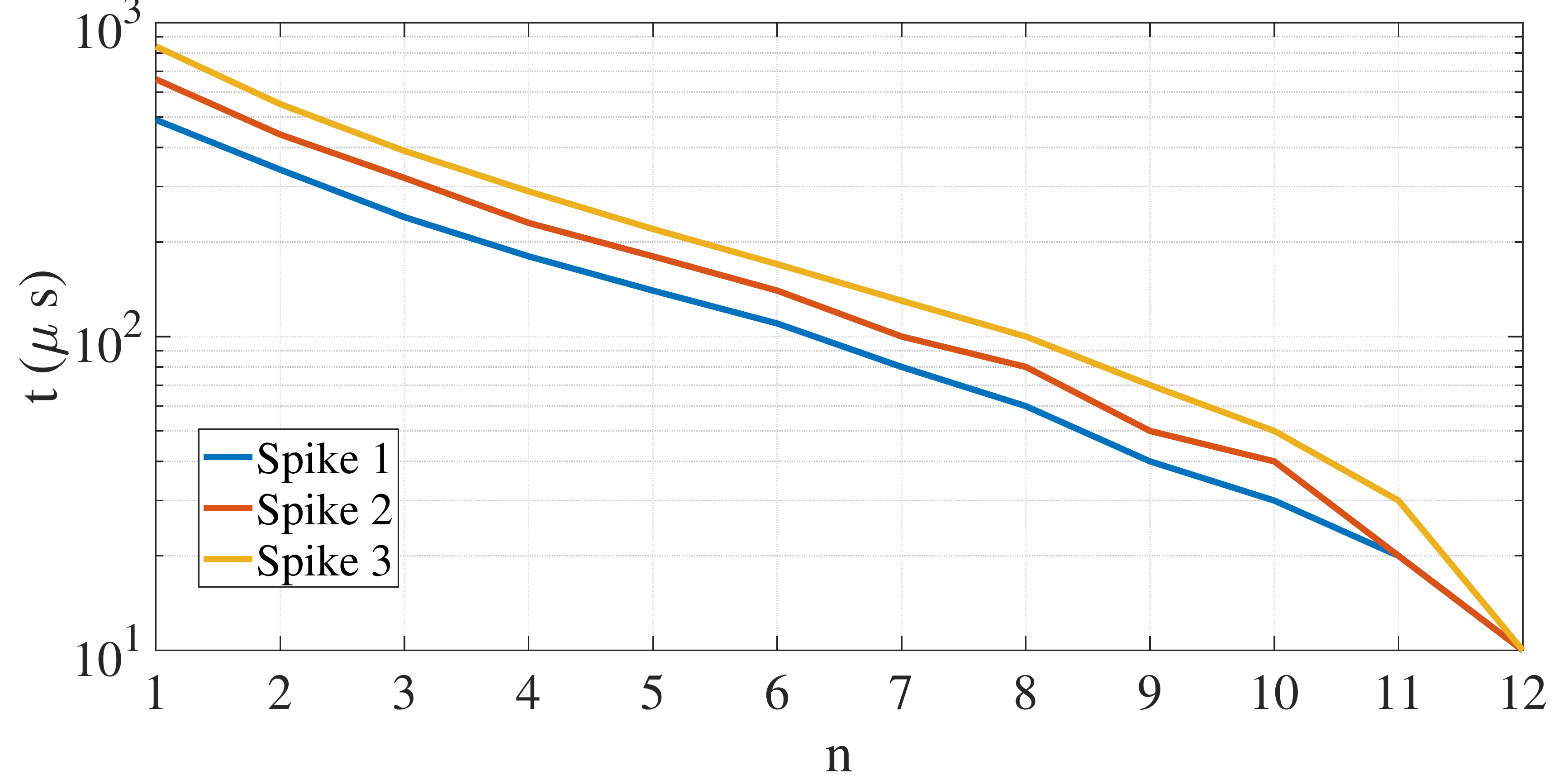}}%
    ~
    \subfigure[][Time offset of spike onsets]{%
    \label{fig:toso}%
    \includegraphics[width=0.4\columnwidth]{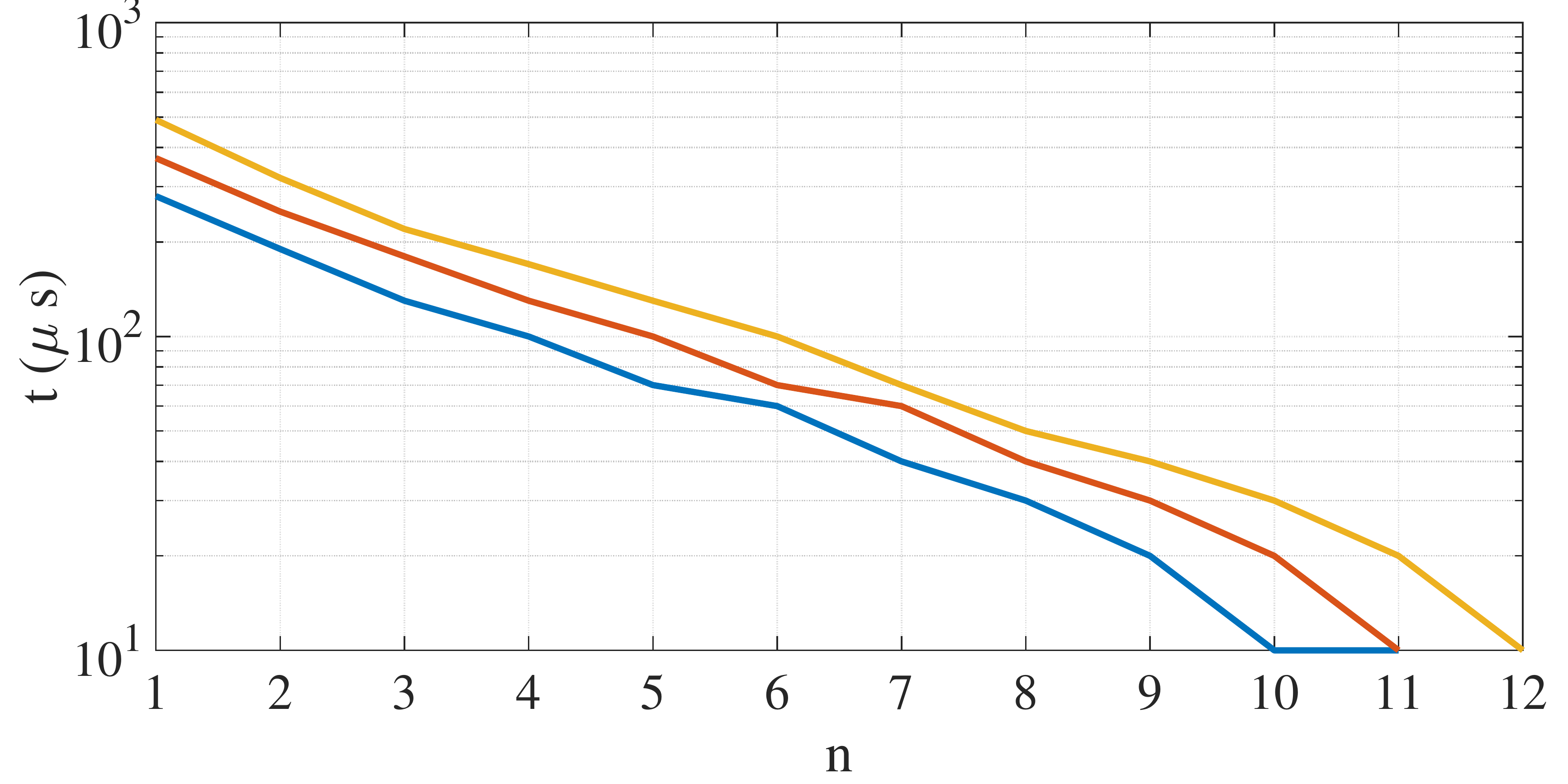}}%
    \\
    \subfigure[][Ratio of spike amplitudes]{%
    \label{fig:pr}%
    \includegraphics[width=0.4\columnwidth]{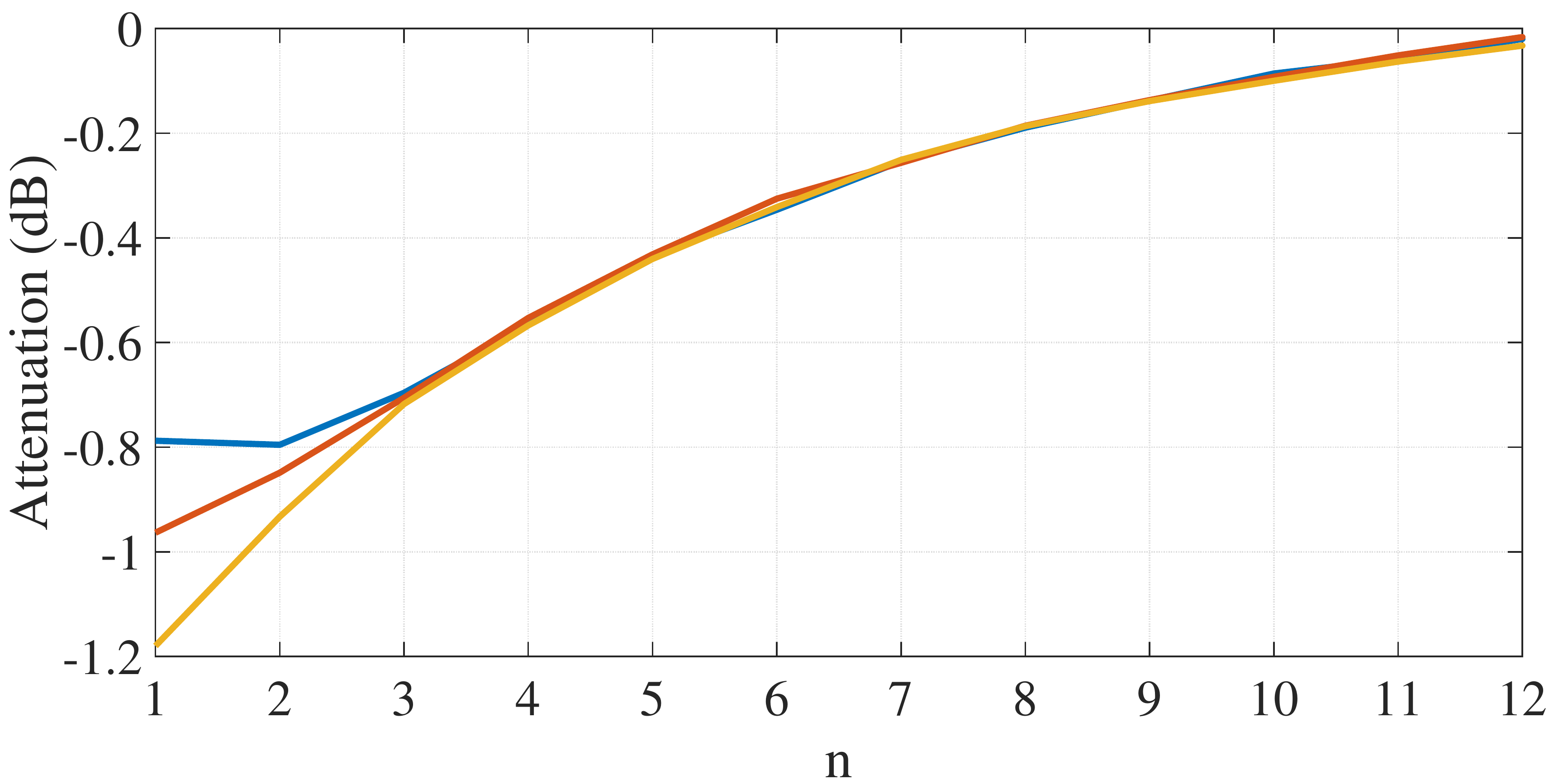}}%
    ~
    \subfigure[][Ratio of pulse widths]{%
    \label{fig:wr}%
    \includegraphics[width=0.4\columnwidth]{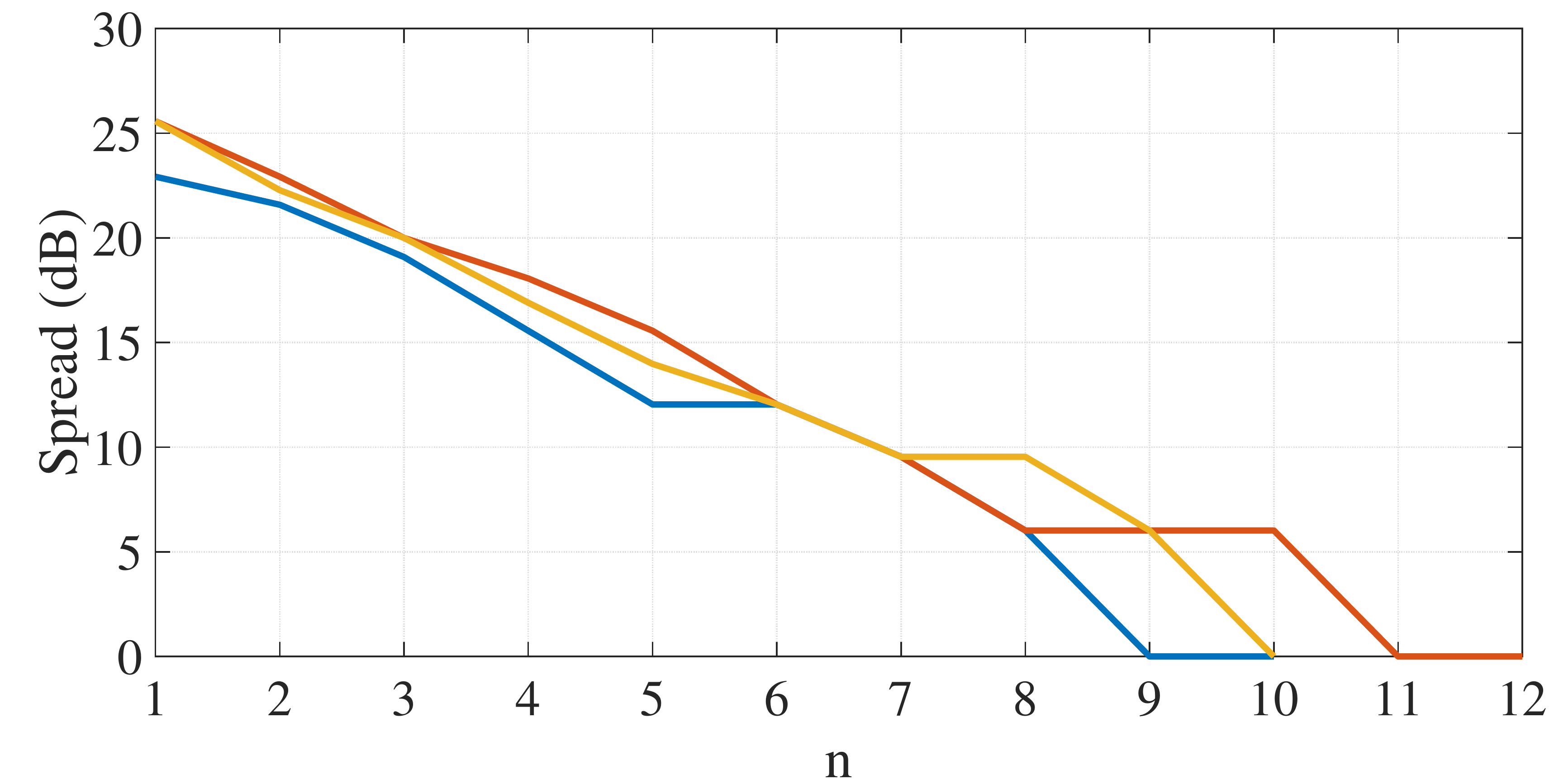}}%
    
    \caption{Exponential relationships between signals from different demyelinated neurons compared to the healthy $n=13$ case.}
    \label{fig:expsig}
\end{figure}

In Figures~\ref{fig:tosp} and \ref{fig:toso} we verify that time intervals represented here correspond to $\Delta t=t_n-t_{13}$ for $1 \le n\le 12$, i.e., they are the lag observed between 13-sheath neuron, which will be the input to our model and other analysed scenarios that the model needs to approximate well, given the value of $n < 13$). 

\begin{figure}[!t]
    \centering
    \subfigure[][Parameter estimation for the transfer function.]{%
    \label{fig:approxim}%
    \includegraphics[width=0.4\columnwidth]{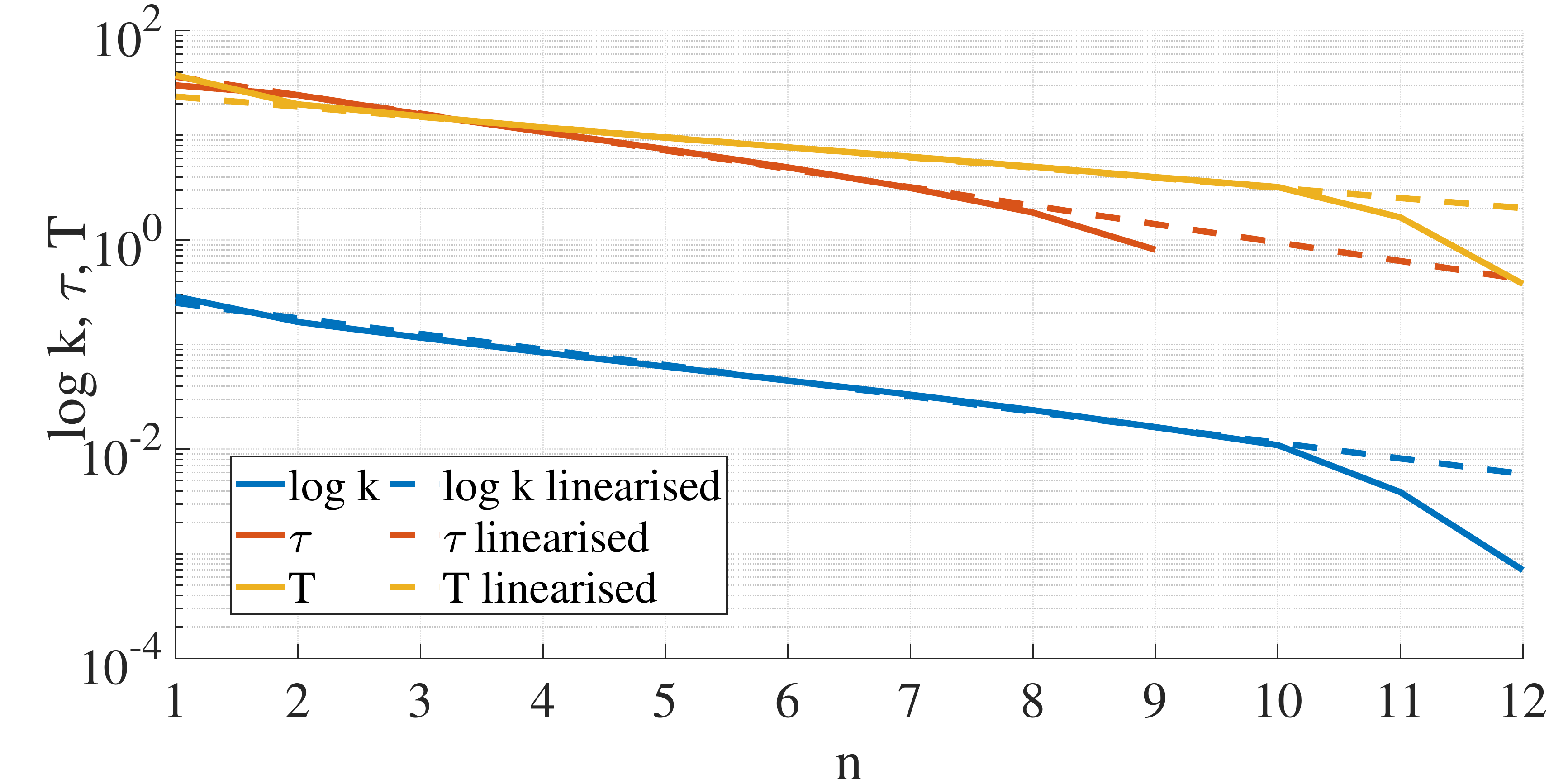}}%
    ~
    \subfigure[][Ratio of RMSE for approximating axonal demyelination.]{%
    \label{fig:rmse}%
    \includegraphics[width=0.4\columnwidth]{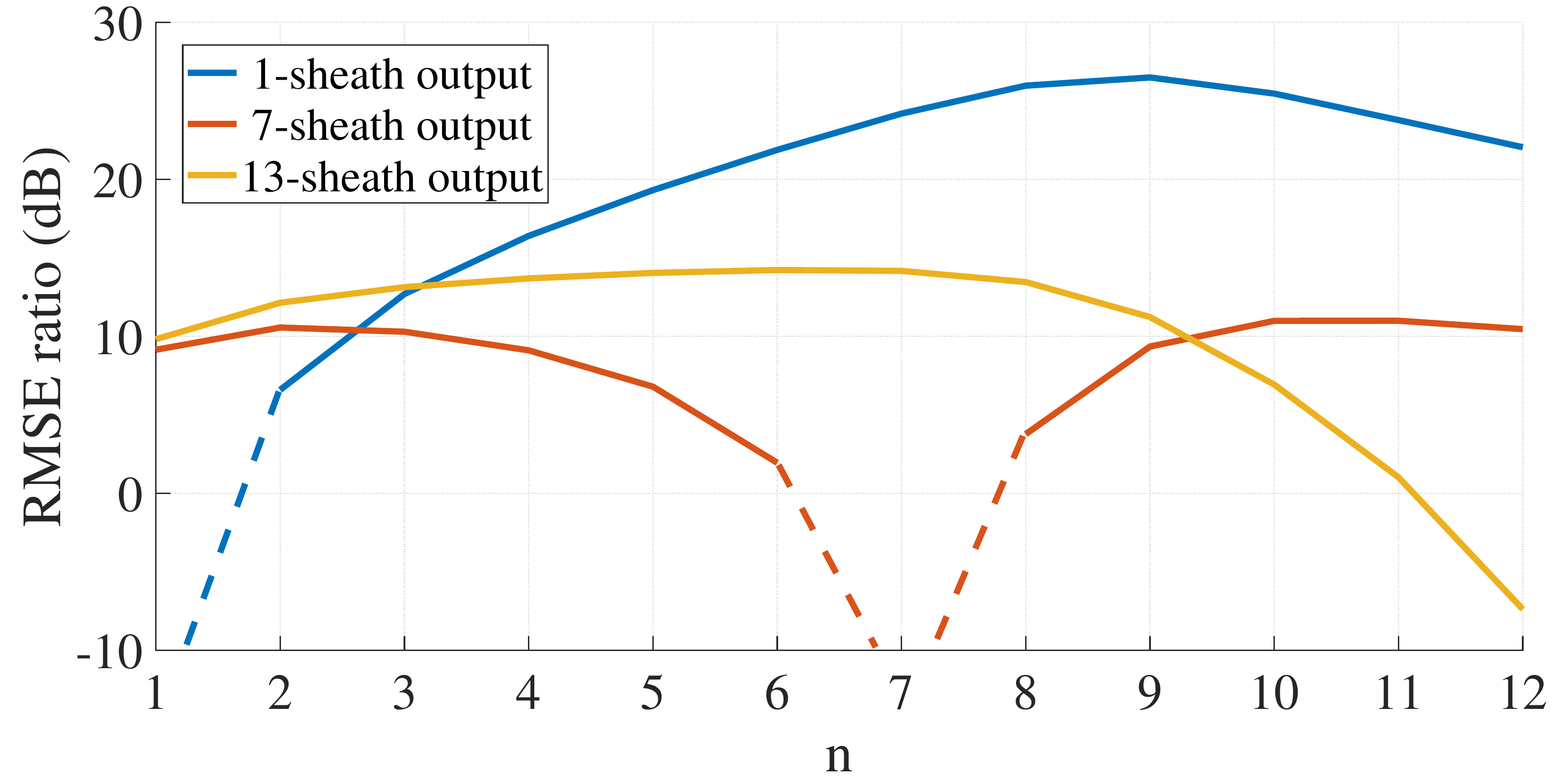}}%
    \caption{Estimation and performance for a linear model of demyelination; (a) Estimated parameters of the transfer function, and their exponential approximation and, (b) Ratio of the RMSE for approximating a demyelinated with the single sheath case, average demyelinated neuron, or a healthy neuron and RMSE for approximating the demyelinated neurons with our transfer function.}
    \label{fig:eight_nine}
\end{figure}

At this point, we can say that (1) for $1\le n\le 10$, we see an exponential decay in the ``lag'' as $n$ grows; (2) spike onsets reach the values observed in $n=13$ one by one ($1^{\text{st}}$ spike for $n=10$, $2^{\text{nd}}$ for $n=11$, 3$^\text{rd}$ for $n=12$) while all the peaks reach the $n=13$ time values in the same case of $n=12$. The first conclusion suggests that we will have a transport delay term $e^{-\tau_n s}$ in the transfer function, in which the delay $\tau_n$ will be an exponential function of $n$. The second conclusion suggests that this term cannot explain all of the dynamics: some of the lag is contributed by a real pole $-1/T_n$, i.e. a term $(1+T_ns)^{-1}$ in the transfer function. Furthermore, after $n=10$, the delay term vanishes and the only effect seen is the one of the pole (we will ignore this effect, as our approximation focus will be for the interval up to $n=10$). Again, given the linearity of the log plot, i.e. exponential nature of the curve, it is expected that $T_n$ is an exponential function of $n$.

In Figures~\ref{fig:pr} and \ref{fig:wr} we observe the behaviour of logarithms of spike amplitudes and pulse widths in the region of interest, which suggests (1) time-invariance of the system, as all three pulses collapse in the same amplitude curve, and (2) that the change in the pulse width requires the real pole $-1/T_n$. This reasoning, graphically presented in Figure~\ref{fig:logic}, confirms our hypothesis about the applicability of the FOPTD transfer function (\ref{eq:tf_FOPTD}) and its exponential coefficients form (\ref{eq:tf_exp}). 

The identified parameters of the model \ref{eq:tf_exp} are $a_{0} = 0.35$, $a_{r} = 0.7$, $\tau_{0} = 54.42$, $\tau_{r} = 0.66$, $T_{0} = 20.27$ and $T_{r} = 0.8$. Those values were found with the Levenberg-Marquardt~\cite{more1978levenberg} numerical optimisation. The iterative procedure was conducted by determining the values of $k_n,\,\tau_n,\,T_n$ for $n=6$, then using those values as initial guesses for $n=5$ and $n=7$, and subsequent values. The relationship between exponential approximation, or linearisation in log domain, and the best choice of coefficients without exponential law assumption is shown in Figure~\ref{fig:approxim}.

It is expected that this would be a good approximation for the observed signals in the ``exponential domain'', $1\le n\le 10$. While the approximation can be accurate outside of this domain as well, we focus on applicability within the range, and we verified it using the RMSE metric (\ref{eq:metrice}) introduced earlier.







Figure~\ref{fig:rmse}, representing $M_n$ for $1\le n\le 12$, gives the answer to the following question: if one ignores the variability in $n$ and replaces every output signal $(x_n)$ with (a) one of a completely deteriorated neuron $N=1$, (b) averagely damaged neuron $N=6$, or (c) a healthy neuron $N=13$, how high is the amplitude of error, compared to that of our model. As expected, for $n=N$ this ratio goes down to $-\infty\text{ dB}$ as the approximation with exact signals is perfect. However, for any other value, even $N\pm1$, our approximation is superior (i.e. above $0$ dB).

\section*{Conclusion}
\label{sec:conclusion}

In this work, we proposed an end-to-end model that takes advantage of the fact that viruses {\color{black}can invade the nervous system} and affect the brain. This model expresses the dynamics of a cytokine storm and its relation to the amount of demyelination that may subject to a neuron. From evidence found in the literature, we developed a model capable of mimicking virus-induced demyelination's degenerative effects. We also proposed a transfer function aiming towards a linear model of the demyelination. Using traditional control and systems theory, we propose a FOPTD function that could help the design of chemical control loops for the reinforcement of myelin. {\color{black}The results show that the demyelination induced by the cytokine storm not only degrades the signal but also compromises the communications among neurons. The signal is attenuated and shifted in time and influences the release of excitatory neurotransmitters into the synaptic cleft. This whole analysis led to the development of a transfer function that fundamentally represents the process of demyelination itself. It not only decreases the level of complexity of the system linking itself with the behaviour of an RC circuit but also underlies the physical rationale of the system by applying biophysically-plausible transfer function models.}

We believe that the proposed models will contribute to bioengineering approaches for neurodegeneration, especially demyelinating disease. For future work, we plan to couple our computational modelling with wet-lab experiments to assess and improve our model's accuracy. Such a technique could pave the way for more sophisticated and precise approaches for the treatment of neurodegeneration. 

\section*{Author Contributions}

GA performed the simulations, data analysis and wrote the first draft of the manuscript. HS performed the linear model analysis. MB and SB led work development. All authors contributed to manuscript writing and revision. All authors have also read and approved the submitted version.

\section*{Acknowledgments}

This publication has emanated from research conducted with the financial support of Science Foundation Ireland (SFI) for the CONNECT Research Centre (13/RC/2077). Figures~\ref{fig:viral_persistence_demyelination} and \ref{fig:long_section_axon} were created with \url{BioRender.com}.

\bibliography{main}




\end{document}